\documentclass[aps,amsmath,amssymb,showpacs,showkeys]{revtex4-2}
\usepackage[dvips]{graphicx}
\usepackage{times}
\usepackage{braket}
\usepackage{xcolor}
\usepackage{adjustbox}
\usepackage{orcidlink}
\usepackage{hyperref}
\hypersetup{
  colorlinks=true,
  urlcolor=blue,
  linkcolor=red,
  citecolor=blue
}

\begin{document}
\title{Strong field lensing and shadows of $\boldsymbol{f(Q,\mathcal{B})}$ 
gravity black holes}

\author{Rahul Das\orcidlink{0009-0008-7579-1229}}
\email[Email: ]{rahulrddas4@gmail.com}

\author{Umananda Dev Goswami\orcidlink{0000-0003-0012-7549}}
\email[Email: ]{umananda@dibru.ac.in}

\affiliation{Department of Physics, Dibrugarh University, Dibrugarh 786004, 
Assam, India}

\begin{abstract}
We investigate the strong field lensing of light by black holes (BHs) 
emerging in $f(Q, \mathcal{B})$ theory of gravity, which is an extension of 
$f(Q)$ gravity theory by including a boundary term $\mathcal{B}$. Our 
analysis starts with the study of horizon structure for the BH spacetime in 
$f(Q,\mathcal{B})$ gravity. The strong field lensing analysis is done by 
employing the method developed by V.~Booza. We calculate the deflection angle 
for the static and spherically symmetric $f(Q,\mathcal{B})$ gravity BH 
spacetime and compare the results with the Schwarzschild BH. It is 
found that for all considerable scenarios of model parameters the 
$f(Q,\mathcal{B})$ gravity BH produces less deflection of light than 
the Schwarzschild BH. We extend our study to examine the outermost 
relativistic Einstein rings and three other observables, viz., angular 
position $\vartheta_{\infty}$, angular separation $s$ and relative 
magnification $r_\text{mag}$. We see that except $r_\text{mag}$, other 
observables show the same kind of behavioural change with respect to model 
parameters. We also study the shadow of the considered BH. Moreover, using 
the shadow data of BH Sgr A* from EHT collaborations, we constrain the model 
parameters of the $f(Q,\mathcal{B})$ theory for the BH spacetime. The 
constraining analysis reveals that for all considerable values of the model 
parameter $s_0$, the allowed range of the other model parameter $q_0$ is 
$-3\lesssim q_0 \lesssim -2$. 
\end{abstract}


\maketitle                                                                      

\section{Introduction}
The theory of General Relativity (GR), put forward by Albert Einstein in 
1915, is a revolutionary theory of gravity, which has predicted the existence 
of gravitational waves (GWs) and black holes (BHs) \cite{1}. Decades after 
the predictions, GWs were discovered by the LIGO and 
Virgo teams in $2015$ \cite{2,3,4,5,6,7}, and the first snap of the direct 
image of the BH at the center of galaxy M$87$ was taken by the Event Horizon 
Telescope (EHT) group in $2019$ \cite{8,9,10,11,12,13}. These recent 
remarkable supports and other past observational validations established GR as 
the most fundamental and robust theory of gravity so far. However, there are 
some issues related to it that need to be addressed so that a complete theory 
of gravity can be obtained. In the high-energy regime, it faces the 
renormalization issue and thus it can not be quantized in a usual manner 
\cite{14}. Moreover, it cannot explain the accelerated expansion of the 
Universe \cite{15, 16} and the flat rotation curve of galaxies \cite{16a,16b} 
without considering the existence of dark components in the Universe, viz., 
dark energy \cite{16c,16d} and dark matter \cite{16e}, respectively. It also has 
compatibility issues with quantum mechanical phenomena \cite{17} as it is 
not a quantum theory of gravity. All of these issues have inspired the 
scientific community to look for other theories of gravity, and to mitigate 
the issues faced by GR, some physicists have been trying to modify the 
geometric part of Einstein's field equations from different perspectives, 
which led to the formulation of modified theories of gravity (MTGs) 
\cite{18,19,20,21,22}. The two most common, widely used MTGs are 
$f(R)$ gravity \cite{18,19,20} and $f(R,T)$ gravity \cite{21,22}. Whereas some 
other physicists have completely changed the way gravity is described, which 
leads to alternative theories of gravity (ATGs). ATGs are based on 
fundamentally different geometrical approaches than the approach of GR. Some 
of the most well known ATGs are TeVeS gravity \cite{23,24}, $f(Q)$ gravity 
\cite{25,26}, $f(T)$ gravity \cite{27}, etc.  

Weyl and Einstein, during their attempt to unify gravity and electromagnetism, 
developed the first theory, where gravity is mediated by torsion \cite{28}, 
rather than by curvature. Many decades later, it was realized that teleparallel 
theories of gravity can also be formulated in flat, torsionless geometries, if 
one attributes gravitational phenomena to the so-called non-metricity tensor 
\cite{29,30}. These three geometries, i.e., the curvature, the torsion and the 
non-metricity, are distinct but provide equivalent descriptions of GR 
(curvature) and all of these theories are rooted in the mathematical 
framework of metric-affine geometry \cite{31,32}. $f(Q)$ gravity, where the 
theory is formulated from an arbitrary function of non-metricity scalar $Q$, 
has played a significant role in addressing cosmological tensions, presenting 
exciting possibilities for reshaping our understanding of gravity and its
 manifestations in cosmology. As $f(Q)$ models have the potential to 
elucidate phenomena in both early and late-time cosmology without 
necessitating the inclusion of dark energy or dark matter \cite{33}, thus 
the $f(Q)$ gravity has gained considerable popularity in the past couple of 
years, and the bulk of the research efforts have been concentrated on 
cosmological applications \cite{34,35,36,37}. This theory has also been 
applied to large structure formation \cite{38}, the development of 
relativistic versions of Modified Newtonian Dynamics (MOND) \cite{39,40}, 
bouncing cosmologies \cite{41,42}, and even in quantum cosmology \cite{43}. 
Extensions like inclusion of boundary terms or nonminimally coupled scalar 
fields were made in order to constrain or test $f(Q)$ models. It was found 
that the boundary term $\mathcal{B}$ included $f(Q)$ gravity, i.e., 
$f(Q,\mathcal{B})$ gravity is dynamically equivalent to $f(R)$ gravity 
\cite{44}. Moreover, there are several relevant studies on the 
$f(Q,\mathcal{B})$ gravity theory with cosmological applications that have 
been investigated in Refs.~\cite{45,46,47}.

Right after the geometric formulation of gravity in GR, the first attempt to 
prove the theory was the studying influence of gravity on the tracks of 
light around a massive object and measuring the bending of photons. Thus, the 
phenomenon of gravitational lensing emerged from GR \cite{48}. In $1919$, 
Eddington, Dyson and Davidson detected the deflection in 
the path of light emerging from stars in the Hyades cluster, around the sun 
during a solar eclipse \cite{49}. A typical lensing system comprises a source, 
a lensing mass and the observer. The rays from the source get deflected by the curvature generated from the mass of the lens and reach the observer. Lensing produces two sets of magnified and distorted images, each on one side of the lens. There are two types of lensing for a specific source and lens system, 
weak lensing \cite{50,50a}, which corresponds to 
the weak gravity application at far distant photons and the strong field 
lensing \cite{51,52}, which corresponds to the application of strong 
gravitational field to the close approaching photons. Particularly, strong 
lensing exhibits distinct and notable features. In such propagation, light 
photons undergo multiple circular loops around the lensing object before 
escaping towards the observer. As a result, an infinite sequence of discrete 
images is formed, which are called relativistic images \cite{53,54}. 
Gravitational lensing can also act as a robust astrophysical tool for the 
study of the gravitational field of massive lensing objects.  

Darwin's foundational contribution \cite{50} marked the beginning of strong 
gravitational lensing studies and established the theoretical basis for later 
investigations. Such studies in the strong field limit have gained 
considerable recognition in recent years, as more information on BHs can be 
extracted from such investigations. Significant advancements in this direction 
were later achieved by Virbhadra and Ellis, who were the first to formulate the 
lens equation \cite{55,56} and analyzed the relativistic images of 
Schwarzschild BHs \cite{57,58}, which later have been extended to naked 
singularities \cite{59,60}. Another aspect of gravitational lensing is the 
time delay in the strong field limit, and various studies \cite{61,62,62a,62b} 
have covered this aspect. Such studies provide valuable insights into 
spacetime geometry and may serve as probes for testing different theories of
gravity and compact object models. Modern studies increasingly consider 
different environments around BHs and investigate their influence on 
the deflection observables. Some of the environments explored earlier are dark 
matter, which influences the geometric curvature of the spacetime \cite{63}; 
inclusion of charged and uncharged plasma \cite{64,65,66}, which alters the 
refractive index of background fields, etc. In fact, in the past decade, 
significant progress has been made in studies of gravitational lensing, 
including investigations for regular BHs and wormholes \cite{67,68,69}. 
Parallel efforts have been made on the studies of strong field deflection in 
the contexts of MTGs and ATGs, revealing remarkable phenomena such as 
relativistic images and shadow formation \cite{70,71}. 
However, for spacetime solutions in MTGs or ATGs, lensing observables show 
deviations from the Schwarzschild or other metrics in GR. 
Analyzing the strong field limit of gravitational lensing in such scenarios, 
one can constrain the parameters of a theory with the help of shadow data 
calculated by the EHT collaboration \cite{72}.

Based on the motivations drawn above, our study aims to explore the strong 
lensing and shadow characteristics of BHs within the framework of the 
$f(Q,\mathcal{B})$ gravity. A BH solution has recently been formulated in this 
gravity in Ref.~\cite{}, and we are interested in exploring lensing aspects 
of this solution as well as its shadow characteristics. The rest of the work is 
structured as follows: In Section \ref{II}, we outline the basic formalism of 
the $f(Q,\mathcal{B})$ gravity and study the horizon structure of the BH 
solution in Section \ref{III}. In Section \ref{IV}, we analyzed the deflection 
angle and located the photon sphere of the BH. In Section \ref{V}, we 
calculate and discuss the behaviour of lensing observables for the considered 
BH spacetime. We compute the shadow radius and constrain the model parameters 
in Section \ref{VI}. Finally, we conclude our work in Section \ref{VII}. 

\section{$\boldsymbol{f(Q,\mathcal{B})}$ Gravity} \label{II}
As mentioned already, the $f(Q)$ gravity \cite{25,26} is an 
alternative geometric formulation of gravity, fundamentally different from 
Einstein's GR. The structure of spacetime in $f(Q)$ gravity is based on the 
symmetric teleparallelism ($R^\alpha_{\beta\mu\nu}=0$) and the non-metricity 
condition ($\nabla_{\beta}g_{\mu\nu} \ne 0$), and hence in this theory 
gravitational phenomena are described in terms of non-metricity scalar $Q$, 
like the Ricci scalar $R$ does in GR, where $R$ represents the amount of 
curvature in spacetime. Thus, $f(Q)$ gravity is a nonlinear extension of the 
symmetric teleparallel theory equivalent to GR (STEGR). The non-metricity 
condition gives non-metricity tensor as given by 
$Q_{\beta\mu\nu} =\nabla_{\beta}g_{\mu\nu}$ and the non-metricity scalar $Q$ 
is obtained as $Q = -\,Q_{\beta\mu\nu}P^{\beta\mu\nu}$. Here, 
$P^{\beta}{}_{\mu\nu}$ is the super potential tensor defined as 
\begin{equation}
P^\beta{}_{\mu\nu} = -\,\frac{1}{2} L^\beta{}_{\mu\nu}
+ \frac{1}{4} \left[\left(Q^\beta - \tilde{Q}^\beta \right) g_{\mu\nu}
- \delta^\beta_{(\mu} Q_{\nu)}\right], \label{eq1}
\end{equation}
where $L^\beta{}_{\mu\nu}$ is the disformation tensor, which is expressed as
\begin{equation}
L^\beta{}_{\mu\nu} = \frac{1}{2} \left( Q^\beta{}_{\mu\nu} - Q_\mu{}^\beta{}_\nu - Q_\nu{}^\beta{}_\mu \right),
\label{eq1a}
\end{equation}
and $Q^{\beta}$ and $\tilde{Q}^{\beta}$ are two characteristic contractions 
of the non-metricity tensor as given by
\begin{equation}
Q_\beta = g^{\mu\nu} Q_{\beta\mu\nu} = Q_{\beta\nu}{}^{\nu},
\quad
\tilde{Q}_\beta = g^{\mu\nu} Q_{\mu\beta\nu} = Q_{\nu\beta}{}^{\nu}. \label{eq2}
\end{equation}
The $f(Q)$ gravity action is 
\begin{equation}
S_{f(Q)} = \int \sqrt{-g}\, d^4 x \left[ \frac{1}{2\kappa^2}f({Q}) + 
\mathcal{L}_m \right], \label{eq3}
\end{equation}
where $\mathcal{L}_m$ is the Lagrangian of the matter field, $g$ is the
determinant of the metric tensor $g_{\mu\nu}$ and ${\kappa}^2 = 8\pi G/c^4$ 
with $G$ as the Newton gravitational constant. The connection coefficient
associated with the theory is 
\begin{equation}
\Gamma^\beta{}_{\mu\nu}=\mathring{\Gamma}^\beta{}_{\mu\nu}+L^\beta{}_{\mu\nu}, 
\label{eq3a}
\end{equation} 
where $\mathring{\Gamma}^\beta{}_{\mu\nu}$ is the Levi-Civita connection.

However, Lorentz non-invariance of the geometric object $Q$ raises some 
serious issues in this theory \cite{26a}. For example, for the coincident 
gauge with static and spherical symmetry, the equations of motion of $f(Q)$ 
gravity indicate a theory with a constant non-metricity scalar or linear 
$f(Q)$ function \cite{26b}. Thus, a correction or an extension in the theory 
needs to be incorporated to account for this required invariance. To proceed 
in this direction, we consider a spacetime with zero curvature (teleparallel 
condition) along zero torsion, from which one can have the following relation: 
\begin{equation}
\mathring{R}^{\beta}{}_{\alpha\mu\nu}
+ \mathring{\nabla}_{\mu} L^{\beta}{}_{\nu\alpha}
- \mathring{\nabla}_{\nu} L^{\beta}{}_{\mu\alpha}
+ L^{\beta}{}_{\mu\rho} L^{\rho}{}_{\nu\alpha}
- L^{\beta}{}_{\nu\rho} L^{\rho}{}_{\mu\alpha}
= 0,  \label{eq4}
\end{equation}
where $\mathring{\nabla}$ and $\mathring{R}^{\beta}{}_{\alpha\mu\nu}$ denote 
the derivative and  the curvature tensor, respectively, corresponding to 
the Levi-Civita connection. Reducing Eq.~\eqref{eq4} after appropriate 
contractions, one can have 
\begin{equation}
\mathring{R} = Q - \mathring{\nabla}_{\beta}\! \left(Q^{\beta} - \tilde{Q}^{\beta} \right). \label{eq5}
\end{equation}
Eq.~\eqref{eq5} shows that the non-metricity scalar alone cannot produce a 
theory equivalent to GR as it differs from the the Ricci scalar by a term of 
total derivative, defined as a boundary term $\mathcal{B}$ as 
\begin{equation}
\mathcal{B} =  \mathring{\nabla}_{\beta}\! \left( Q^{\beta} - \tilde{Q}^{\beta} \right). 
\label{eq6}
\end{equation}
Thus, keeping in view of this, it is essential to extend or modify the $f(Q)$
gravity theory by including the boundary term $\mathcal{B}$ in it to have a 
STEGR that has the required Lorentz invariance, and the resulting theory is 
termed as the $f(Q,\mathcal{B})$ gravity theory \cite{73,74}, whose the 
action can be written straightforwardly as 
\begin{equation}
S_{f(Q,\mathcal{B})} = \int \sqrt{-g}\, d^4x \left[\frac{1}{2\kappa^2} f(Q,\mathcal{B}) + \mathcal{L}_m \right]. \label{eq7}
\end{equation}
The equations of motion for the $f(Q,\mathcal{B})$ gravity can be found by 
varying the action \eqref{eq7} with respect to the metric tensor, which are 
obtained as
\begin{align}
f_Q(Q,\mathcal{B})\,\mathring{G}_{\mu\nu} - 
\frac{1}{2}\, g_{\mu\nu} \big[f(Q,\mathcal{B}) - 
f_Q(Q,\mathcal{B})\,Q & - f_\mathcal{B}(Q,\mathcal{B})\mathcal{B} \big] + 
2 P^{\alpha}{}_{\mu\nu} \partial_{\alpha} \big[ f_Q(Q,\mathcal{B}) + 
f_\mathcal{B}(Q,\mathcal{B}) \big] \notag\\[5pt]
& - g_{\mu\nu} \mathring{\Box} f_\mathcal{B}(Q,\mathcal{B})
+ \mathring{\nabla}_{\mu}\mathring{\nabla}_{\nu} f_\mathcal{B}(Q,\mathcal{B})
= \kappa^2\, T_{\mu\nu}, \label{eq8}
\end{align}
where $f_Q(Q,\mathcal{B})$ and $f_\mathcal{B}(Q,\mathcal{B})$ represent the 
derivative of $f(Q,\mathcal{B})$ with respect to $Q$ and $\mathcal{B}$, 
respectively. $\mathring{G}_{\mu\nu}$ is the Einstein 
tensor and $ T_{\mu\nu}$ is the energy-momentum tensor. The addition of the 
boundary term leads to the existence of fourth-order field equations \cite{73}. 
It is obvious that on substitution of $\mathcal{B} = 0$, Eq.~\eqref{eq8} 
reduces to the field equations of $f(Q)$ gravity theory \cite{75,76,77}. As 
mentioned earlier, an interesting property of this theory is that from this
theory the $f(R)$ gravity can be recovered by considering 
$\mathring{R}\equiv R = Q - \mathcal{B}$, i.e., 
%
$f(R) \equiv f(Q,\mathcal{B}) = f(Q-\mathcal{B})$.

\section{Black hole solution in $\boldsymbol{f( Q,\mathcal{B})}$ gravity} 
\label{III}
In this work, we study the gravitational lensing and shadow of a BH solution 
obtained by Ref.~\cite{78} in the $f(Q,\mathcal{B})$ gravity theory with a 
static and spherically symmetric spacetime metric given as  
\begin{equation}
ds^2 = A(r)\,dt^2 - \frac{1}{A(r)}\,dr^2 - B(r)\left(d\theta^2 + \sin^2\theta\, d\phi^2\right), \label{eq10}
\end{equation}
where $A(r)$ and $B(r)$ are metric functions, which are functions of the 
radial coordinate only. This 
metric \eqref{eq10} can be used to evaluate the non-metricity scalar and the 
boundary term in terms of the metric functions $A(r)$ and $B(r)$. The 
non-metricity scalar $Q$ for this metric takes the form:
\begin{equation}
Q(r) = -\, \frac{B'(r)\big(\sqrt{B(r)}A'(r) +2 A(r)B'(r)\big)}{B(r)}, 
\label{eq12}  
\end{equation}
and the boundary term given by Eq.~\eqref{eq6} becomes
\begin{equation}
\mathcal{B}(r) = A''(r) + \frac{3 A'(r)\,B'(r) + 2 A(r)\,B''(r)}{\sqrt{B(r)}} + \frac{16 A(r)\,B'^2(r) - 2}{\sqrt{B(r)}}. \label{eq11}
\end{equation}

\begin{figure}[!h]
       \centerline{
        \includegraphics[scale = 0.55]{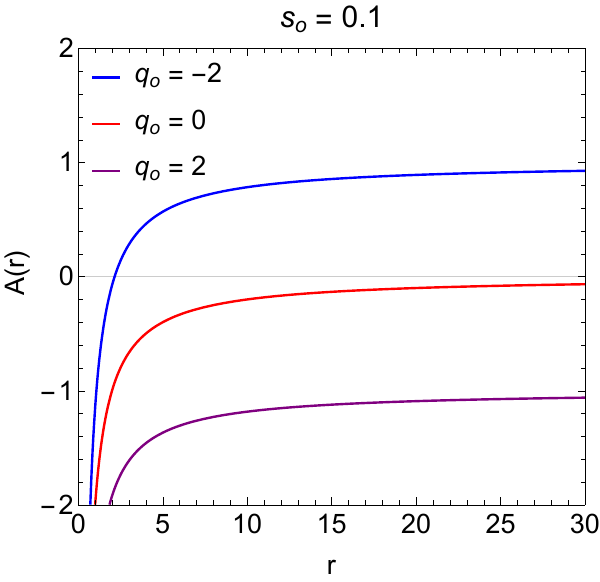}\hspace{0.2cm}
        \includegraphics[scale = 0.55]{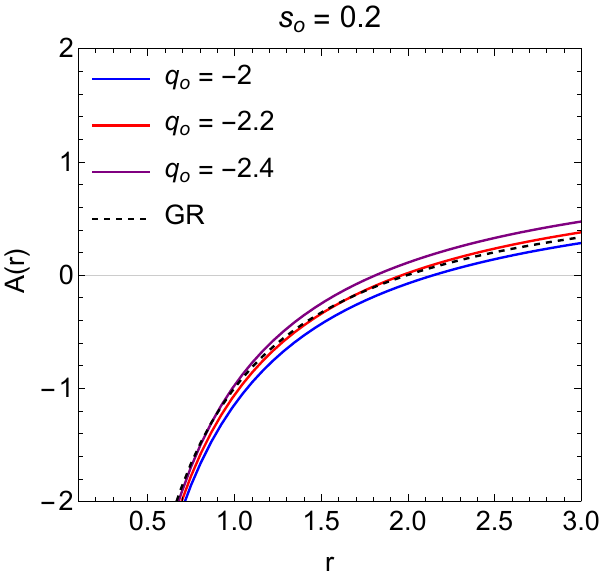}\hspace{0.125cm}
        \includegraphics[scale = 0.55]{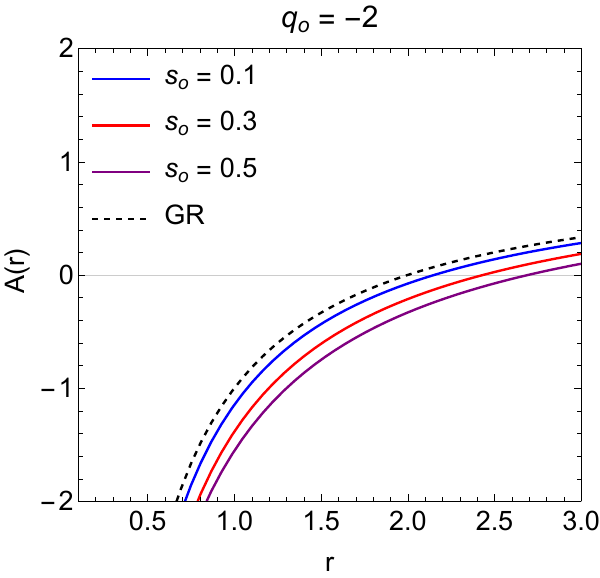}}
        \vspace{-0.2cm}
\caption{Horizon structure of the $f(Q,\mathcal{B})$ gravity BHs for different
values of the model parameters.}
\label{fig1}
\end{figure}

To obtain BH solutions that preserve the Lorentz symmetry, the authors of 
Ref.~\cite{78} have adopted the condition $f_Q = -\,f_\mathcal{B}$. This is a 
necessary condition in the $f(Q,\mathcal{B})$ theory for the coincident gauge 
with static spherically symmetry. Also, for simplicity, it is considered that 
$B(r) = r^2$. As it is very difficult to solve the field equations in 
\eqref{eq8} of $f(Q,\mathcal{B})$ 
gravity for the metric \eqref{eq10}, the authors have adopted a simplified
approach to find an expression for the metric function $A(r)$ from the models
of boundary term and the non-metricity scalar. Accordingly, the solution of
our interest can be obtained from a model of the non-metricity scalar as 
given by
\begin{equation}
Q(r) = \frac{q_0}{{r^2 + {s}^2_0}} \label{eq13},
\end{equation}
where $q_0$ and $s_0$ are two arbitrary constants. Equating this relation 
with that of Eq.~\eqref{eq12} and solving the resulting equation, one can 
obtain the expression of $A(r)$ as \cite{78}
\begin{equation}
A(r) = -\frac{2M}{r} - \frac{q_0}{2} \left[ 1 - \frac{s_0}{r} \tan^{-1}\!\left(\frac{r}{s_0}\right) \right] \label{eq14}.
\end{equation}
This metric function \eqref{eq14} is asymptomatically flat at 
$r\rightarrow \infty$, but diverges as $r\rightarrow 0$. One can recover the 
Schwarzschild solution by setting $q_0 = -\,2$ and $s_0 = 0$ \cite{78}. 
Fig.~\ref{fig1} shows the variation of metric function $A(r)$ with respect to 
the radial coordinate $r$, and it is seen that its behavior is different for 
different $q_0$ and $s_0$ values. From the left plot of this figure, which is
for the fixed value of $s_0=0.1$, we observe that when $ q_0 = 0$, $A(r)$ tends 
to zero with a gradually increasing  value. For $q_0 > 0$, $A(r)$ tends 
towards a fixed negative value and has no horizon. Conversely, for $ q_0 < 0$, 
there is a single horizon and $A(r)$ tends to have a positive value. The 
asymptomatic fixed value of $A(r)$ for positive, zero and negative $q_0$ can be 
found from Eq.~\eqref{eq14}, which shows that $A(r)$ tends to $-\,q_0/2$ in 
the limit $r\rightarrow \infty$. The middle plot of Fig.~\ref{fig1} analyses 
the behavior of $A(r)$ for different negative $q_0$ values with a fixed 
$s_0=0.2$ value. The horizon radius is seen decreasing for increasing negative 
$q_0$ values, with the horizon radius varying very slightly from that of the 
Schwarzschild BH. Similarly, in the right plot of Fig.~\ref{fig1}, we show the 
variation of $A(r)$ for different $s_0$ values and keep $q_0$ fixed at $-2$. 
We can see that the horizon radius increases with increasing the variable $s_0$ 
with values exceeding the Schwarzschild horizon radius.

\begin{figure}[!h]
        \centerline{
        \includegraphics[scale = 0.65]{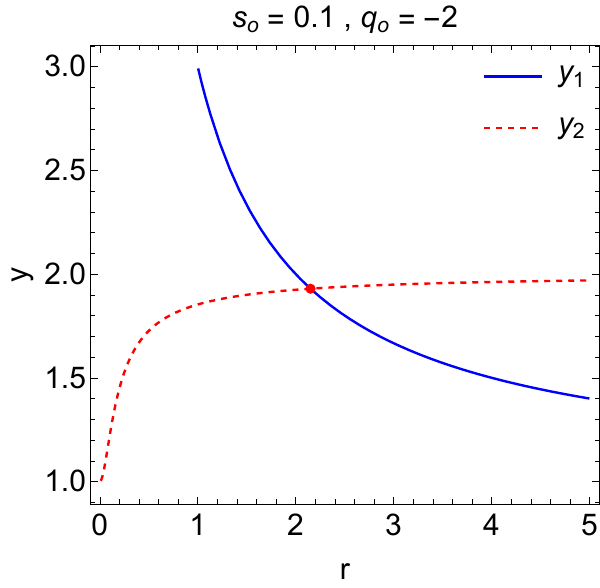}\hspace{0.5cm}
        \includegraphics[scale = 0.65]{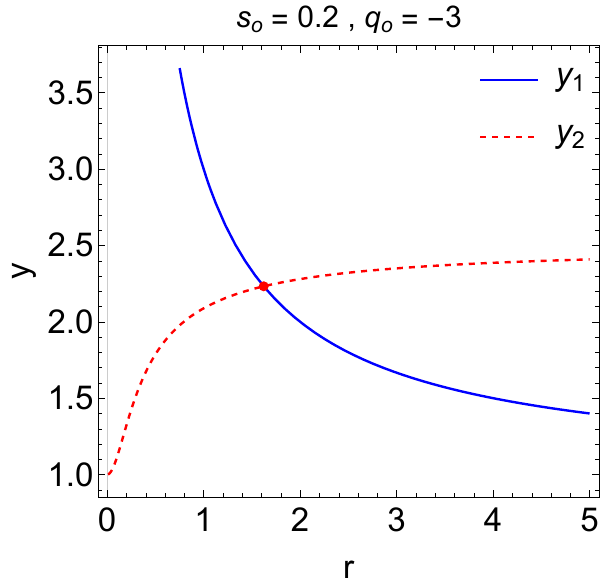}}
        \vspace{-0.2cm}
\caption{Graphical method of locating the BH horizon. $y_1 =  1 + 2M/r$ is 
shown by blue line and $y_2 =  1 - q_0/2 \left[1 - s_0/r \tan^{-1}\!\left(r/s_0\right)\right]$ is shown by red dashed line. The red dot denotes the location 
of the horizon.}
\label{fig2}
\end{figure}

As we see in Fig.~\ref{fig1}, the $f(Q,\mathcal{B})$ gravity BH supports 
a single horizon only, and its location can be found from a simple geographic 
method. Since the positive roots of the metric function $A(r) = 0$ correspond 
to the locations of BH horizons \cite{79,80}, we set $A(r) = 0$ in 
Eq.~\eqref{eq14} and rearrange the resulting equation into two parts as 
\begin{center}\begin{equation}
 1 + \frac{2M}{r} = 1 - \frac{q_0}{2} \left[ 1 - \frac{s_0}{r} \tan^{-1}\!\left(\frac{r}{s_0}\right) \right] \label{eq15}.
\end{equation}\end{center}
The graphical method includes plotting both sides of this Eq.~\eqref{eq15} 
for different values of $q_0$ and $s_0$. In Fig.~\ref{fig2}, we plot both 
sides of Eq.~\eqref{eq15}. In this figure, the blue colored curve represents 
the LHS and the dashed red colored curve represents the RHS of 
Eq.~\eqref{eq15}. As expected, the plots intersect only at one point. The 
intersection of the curves, i.e., the red dot in Fig.~\ref{fig2}, locates the 
horizon radius of the $f(Q,\mathcal{B})$ gravity BHs for the considered 
values of the parameters. From Fig.~\ref{fig2}, it 
is clear that on increasing $s_0$ and decreasing $q_0$, the horizon of the BHs 
shrinks.

\section{Bending of Light in the Strong Field Limit} \label{IV}
In this section, we explore the strong field limit of the deflection of light 
in the $f(Q,\mathcal{B})$ gravity BH spacetime. The strong gravitational field 
applies to those photons that are travelling very close to the BH event 
horizon. These photons are deviated from their usual path and are forced to 
move in bent trajectories. Deflection angle is defined as the amount of 
bending of the path of photons or light from their original path. The 
deflection angle reaches $2\pi$ for a certain minimum value of closest 
approach distance $r_0$, and after that, the deflection angle exceeds more than 
$2\pi$, which means that photons encircle the BH more than once. At the 
critical value $r_0 = r_p$, the deflection angle diverges \cite{51,79,81}. At 
this point, the photons can not escape the potential of the BH spacetime and 
remain circling the BH forever. This minimum value of the closest 
approach distance $r_p$ is called the photon sphere radius. We aim to analyse 
the deflection angle and other related observables for the aforementioned 
$f(Q,\mathcal{B})$ BH solution. We treat the strong lensing phenomenon here 
via the method developed by V.~Bozza \cite{51}. Following this method, we 
consider that the photon trajectories confined in the equatorial plane of the 
static and spherically symmetric BH only. It should be noted that the entire 
trajectory of the photon is governed by the equation 
$g_{\mu\nu}U^{\mu}U^{\nu} = 0$, where $U^{\mu} = dx^{\mu}/d\tau$ with $\tau$ as 
the affine parameter. As mentioned before, we consider $B(r) = r^2$ for 
simplification. The light trajectory equation is then written as \cite{82,83}
\begin{equation}
A(r)\,\dot t^2 - \frac{1}{A(r)}\,\dot r^2 - r^2 \dot \phi^2 = 0.  \label{eq16}
\end{equation}  
For a spherically symmetric spacetime, conservation of energy 
$E = A(r)\dot{t}$ and angular momentum $L = r^2\dot \phi$ follow 
through the journey of photons and define the impact parameter as 
$\zeta = L/E$. The various possible orbits of photons are governed by 
Eq.~\eqref{eq16}, as this null geodesic equation can be rewritten into a form 
analogous to a classical equation of particle motion \cite{82,84}:
\begin{equation}\label{eqo}
\dot r^2 = V_\text{eff}(r),
\end{equation}
where $V_\text{eff}$ is the effective radial potential in the BH spacetime and
can be written as 
\begin{equation}
V_\text{eff}(r) = E^2\left[1-\zeta^2 \frac{A(r)}{r^2}\right]\!. \label{eq17}
\end{equation}
\begin{figure}[!h]
        \centerline{
        \includegraphics[scale = 0.65]{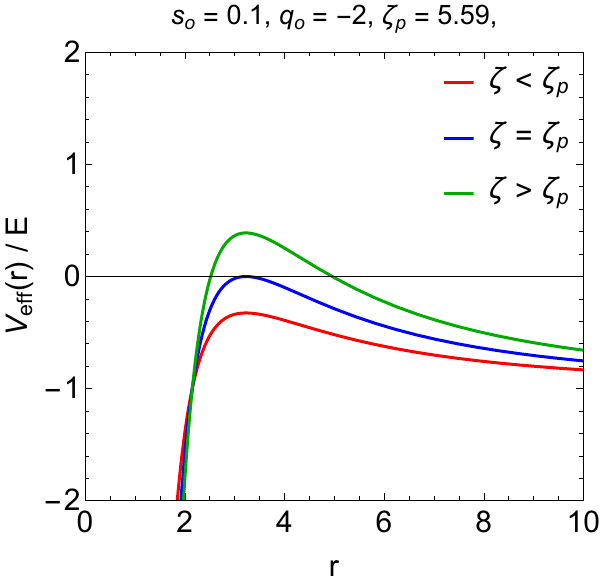} \hspace{0.5cm}
        \includegraphics[scale = 0.65]{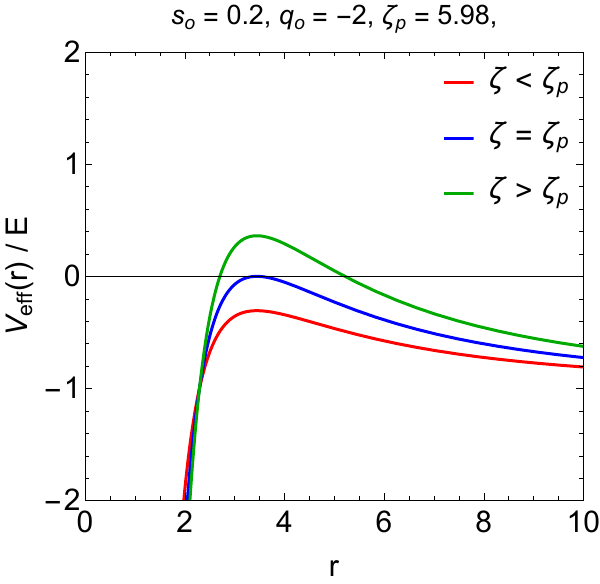}}
        \vspace{-0.2cm}
\caption{Variation of effective potential $V_\text{eff}$ defined in 
Eq.~\eqref{eq17} with respect to radial coordinate $r$ for different values
of impact parameter $\zeta$ and two sets of model parameters.}
\label{fig3}
\end{figure}

\begin{figure}[!h]
        \centerline{
        \includegraphics[scale = 0.65]{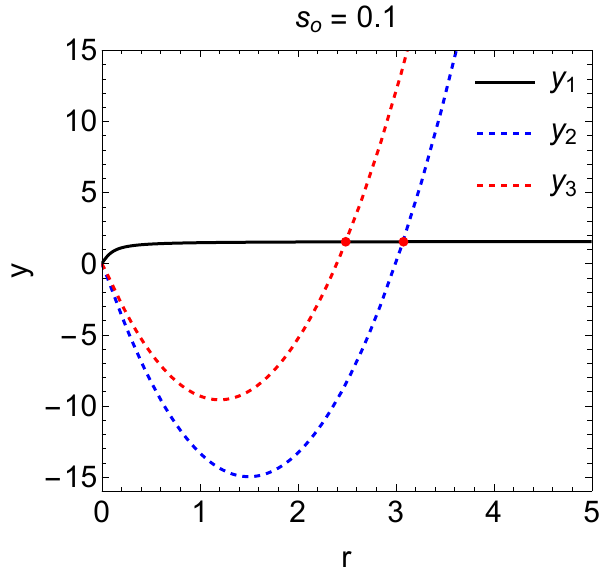}
\hspace{0.5cm}
        \includegraphics[scale = 0.65]{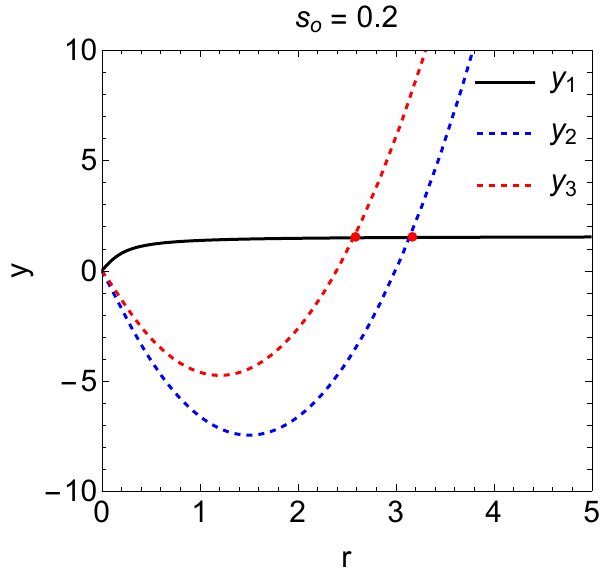}}
        \vspace{-0.2cm}
\caption{Graphical method of locating photon sphere radius from 
Eq.~\eqref{eq20}. Here, in the plots, the solid line ($y_1$) represents 
$\tan^{-1}(r/s_0)$ and dotted lines ($y_2$, $y_3$) represent 
$r/3s_0[(1+ r^2/s_0^2)^{-1} + 2\, r(1+6M/_0r)]$ for two
values of $q_0$, viz., $q_0 = -\, 2$ ($y_2$) and $q_0 = -\, 2.5$ ($y_3$). For 
the left plot, $s_0= 0.1$ and for the right plot $s_0= 0.2$ are used.}
\label{fig4}
\end{figure}

\begin{figure}[!h]
        \centerline{
        \includegraphics[scale = 0.66]{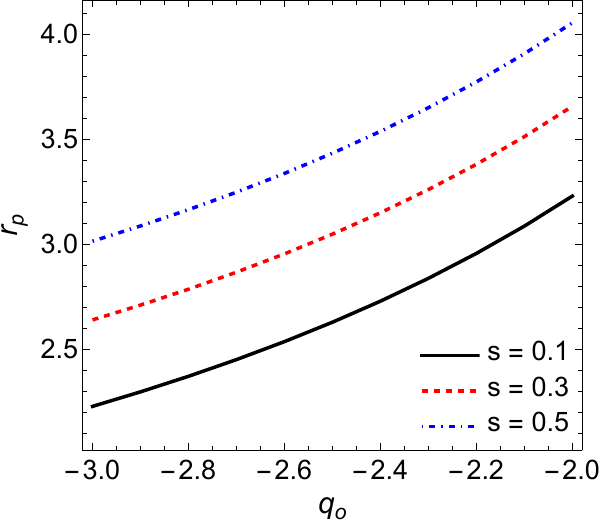}
\hspace{0.5cm}
        \includegraphics[scale = 0.65]{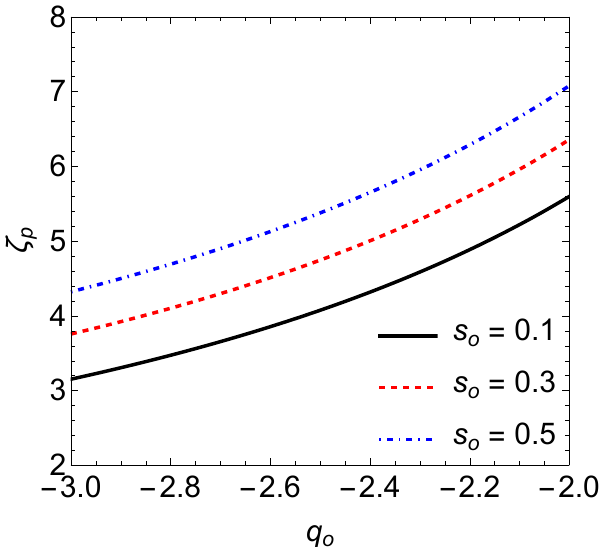}}
        \vspace{-0.2cm}
\caption{Behaviors of photon sphere radius $r_p$ and the corresponding 
critical impact parameter $\zeta_p$ with the parameters $q_0$ and $s_0$.}
\label{fig5}
\end{figure}

Fig.~\ref{fig3} shows the variation of effective potential as a function of the radial coordinate $r$ for two different sets of parameters $q_0$ and $s_0$ and three different values of impact parameter $\zeta$, viz.~$\zeta<\zeta_p$, $\zeta=\zeta_p$ and $\zeta>\zeta_p$, where $\zeta_p$ is the minimum or critical value of impact parameter at the photon sphere radius $r_p$, at which $d V_\text{eff}/dr = 0$ \cite{71,85}. A photon with $\zeta>\zeta_p$ is deflected by the BH while one with $\zeta<\zeta_p$ is captured by the BH \cite{86}. The photon sphere radius is an immediate consequence of Eq.~\eqref{eq17}, where $d V_\text{eff}/dr = 0$ and hence is the largest 
positive root of this equation, i.e.,
\begin{equation}
\frac{d}{dr}\!\left(\frac{A(r)}{B(r)}\right) = 0, \label{eq18}
\end{equation}
which leads to the relation \cite{51,56}:
\begin{equation}
\frac{B'(r)}{B(r)} = \frac{A'(r)}{A(r)} \label{eq19}
\end{equation}
and inserting the metric function given by Eq.~\eqref{eq14}, this relation 
gives an equation as
\begin{equation}
\tan^{-1}\!\left(\frac{r}{s_0}\right) - \frac{r}{3s_0}\left[\left(1+ \frac{r^2}{s_0^2}\right)^{-1}\!\!\! + 2\, r\!\left(1+\frac{6M}{q_0r}\right)\right] = 0.
\label{eq20}
\end{equation}
Clearly, Eq.~\eqref{eq20} can not be solved analytically due to the complexity 
in its structure, and hence needs to be solved numerically to obtain the photon 
sphere radius $r_p$ for different values of parameters $q_0$ and $s_0$. 
Further, for illustrating the locations of the $r_p$ for the specific 
parameter values one can use the graphical method, wherein the first part of 
Eq.~\eqref{eq20}, i.e., $\tan^{-1}\!(r/s_0)$, and the second part 
(within the square brackets) of it, i.e., $r/3s_0[(1+ r^2/s_0^2)^{-1} 
+ 2\, r(1+6M/q_0r)]$ are plotted independently with respect to $r$. As an 
example, Fig.~\ref{fig4} shows such analysis plots for $s_0 = 0.1$ and
$0.2$ with $q_0 = -2$ and $-2.5$. It is seen from each plot of the figure that 
both curves, i.e., the solid curve representing the first part and the dashed 
curve specifying the second part for a given value of $q_0$, intersect at a 
single real positive point (except at the origin), which locates the photon 
sphere radius for the given specific values of $s_0$ and $q_0$. Each plot of 
Fig.~\ref{fig4} indicates that for a fixed value of $s_0$, $r_p$ decreases with 
decreasing the values of $q_0$. Whereas for a given value of $q_0$, $r_p$ 
increases with increasing $s_0$. These dependencies of $r_p$ on $s_0$ and
$q_0$ are clearly shown in the left plot of Fig.~\ref{fig5} as obtained from 
the numerical solution. It should be noted that the photon sphere radius $r_p$ 
corresponds to the peak of the effective potential $V_\text{eff}$ for the 
given values of $s_0$ and $q_0$ (see Fig.~\ref{fig3}), which lies beyond the 
event horizon $r_h$ (see Fig.~\ref{fig1}). Moreover, from the equation of 
orbit \eqref{eqo} of photons around the BH, the impact parameter $\zeta$ can 
be obtained as
\begin{equation}
\zeta(r) = \sqrt{\frac{B(r)}{A(r)}}. \label{eq21}
\end{equation}
Thus, the critical impact parameter $\zeta_p$ at the photon sphere radius,
i.e., at $r = r_p$ can be calculated as 
\begin{equation}
\zeta_ p = \sqrt{\frac{B(r_p)}{A(r_p)}} = \frac{r_p}{\sqrt{ -\frac{2M}{r_p} - \frac{q_0}{2} \left[1 - \frac{s_0}{r_p} \tan^{-1}\!\left(\frac{r_p}{s_0}\right)\right]}}. \label{eq22}
\end{equation}

For more clarity of the picture, we analyzed the behavior of photon sphere 
radius $r_p$ along with the corresponding critical impact parameter $\zeta_p$ 
for different values of parameters $q_0$ and $s_0$ in Fig.~\ref{fig5}. The left 
plot of this figure shows the variation of photon sphere radius as a function 
of $q_0$ over the range of $(-3,-2)$, for three fixed $s_0$ values. We 
can see that the photon sphere radius decreases with increasing the negative 
value or decreasing value of $q_0$, as already mentioned. Similarly, the right 
plot of the figure shows the variation of the critical impact parameter 
$\zeta_p$ as a function of $q_0$. As in the case of $r_p$, this plot also 
shows that the critical impact parameter decreases with decreasing value of 
$q_0$ with fixed $s_0$ value. Further, for increasing values of $s_0$, both
$r_p$ and $\zeta_p$ increase. 

\subsection{Deflection angle in $\boldsymbol{f(Q,\mathcal{B})}$ gravity black
hole spacetime}
The deflection angle $\hat{\alpha}$ produced by the bending of photon 
trajectories can be expressed as \cite{55,87} 
\begin{equation}
\hat{\alpha} = \int\! \left(\frac{d\phi}{dr}\right) dr -\pi = I(r_0)-\pi, \label{eq23}
\end{equation}
where the integrand $I(r_0)$ can be found from the equation of orbit 
\eqref{eq16} as \cite{51,87} 
\begin{equation}
I(r_0) = 2\! \int_{r_0}^{\infty}\!\! 
\frac{1}{\sqrt{A(r)\,B(r)} \sqrt{\frac{\,B(r)A(r_0)}{\,B(r_0)A(r)} - 1}} \, dr. \label{eq24}
\end{equation}
The above integral counts the contributions from the whole trajectory from 
source to observer and treats the contributions for to and from the lens as 
equal. We analyse the strong field deflection angle only for those photons that 
approach very close to the photon sphere of the BH. In order to deal with the 
extremely close incoming photons, we use the method given by Bozza \cite{51}. 
With this method one can evaluate the diverging deflection angle corresponding 
to the critical impact parameter $\zeta_p$ at photon sphere radius $r_{p}$. To 
proceed in this direction, we introduce a new variable, defined as \cite{51} 
\begin{equation}
z = \frac{A(r)-A(r_0)}{1-A(r_0)}. \label{eq25}
\end{equation}
With this new variable given by Eq. \eqref{eq25}, one should be cautious while 
treating all the functions. Those with subscript ``0" to be evaluated at 
$r_0$ and all without this to be evaluated at $r = A^{-1}\left[(1 -A_0 ) z + 
A_0\right]$. Using this new variable the integral $I(r_0)$ can be expressed as 
\cite{51,87,91} 
\begin{equation}
I(r_0) = \int_{0}^{1}\!\!R(z,r_0) \,f(z,r_0)\, dr, \label{eq26}
\end{equation}
where the functions $R(z, r_0)$ and $f(z, r_0)$ are defined as \cite{51,87} 
\begin{align} 
R(z,r_0) & = \frac{2(1-A_0)\sqrt{B_0}}{BA'}, \\[5pt]
f(z,r_0) & = \frac{1}{\sqrt{A_0 -[(1 - A_0)z + A_0]B_0/B}}.
\label{eq28}
\end{align}
The function $R(z,r_0)$ is regular for all values of $z$ and $r_0$, while 
$f(z,r_0)$ diverges at $z\rightarrow 0$. To find the order of divergence of 
the integrand, we expand the argument with the square root in $f(z, r_0)$ to 
the second order in $z$, which gives,
\begin{equation}
f(z,r_0) \equiv f_0(z,r_0) = \frac{1}{\sqrt{\alpha z +\beta z^2}},\label{eq29}
\end{equation}
where 
\begin{align}
\alpha & = \frac{1-A_0}{B_0A'_0}
\left(B'_0A_0-B_0A'_0\right),
\label{eq30}\\[5pt]
\beta & =
\frac{(1-A_0)^2}{2B_0^2 {A'_0}^3}
\left[2B_0B'_0{A'_0}^2 + (B_0B''_0-2{B'_0}^2)A_0A'_0-B_0B'_0A_0A''_0\right].
\label{eq31}
\end{align}
The integral \eqref{eq26} gives a finite result when $\alpha $ is non-zero, as 
in this case the leading order of divergence in $f_0$ is $\,z^{-1/2}\!$, which
can be integrated out. At the photon sphere radius, i.e., when $r_0 = r_{p}$,  
$\alpha$ of Eq.~\eqref{eq30} vanishes and then the leading order of divergence 
is $z^{-1}$, which makes the integral to diverge \cite{51}. Thus, to solve the 
integral, we split it into two parts. One part of it contains the divergence, 
and is given as 
\begin{equation}
I_D(r_0) = \int_{0}^{1}\!\!R(0,r_0)f_0(z,r_0)\,dz, \label{eq32}
\end{equation}
and the other part contains all the regular contributions from the rest of the 
trajectory, other than the near region of the photon sphere, and is given as 
\begin{equation}
I_R(r_0) = \int_{0}^{1}\!\!g(z,r_0)\,dz,\label{eq33}
\end{equation}
where $g(z,r_0) = R(z,r_0)f(z, r_0) - R(0, r_0)f_0(z, r_0)$. 
It is necessary to compute both the integrals separately and combine them to 
get the deflection angle. Further analysis of the integrals, including their 
approximation up to the leading order, produces the deflection angle formula 
in the strong field limit in terms of the impact parameter near its critical 
value, as given by \cite{51}
\begin{equation}
\hat{\alpha}(\zeta) = -\,\bar{a}\,\log\!\left(\frac{\zeta}{\zeta_p} - 1\right)
+ \bar{b}, \label{eq34}
\end{equation}
where $\bar a$ and $\bar b$ are deflection coefficients given by \cite{51}
\begin{align}
\bar{a} & = \frac{R(0, r_{p})}{2 \sqrt{\beta_p}}
= \frac{2\sqrt{2}}{\sqrt{-q_0}}
\left(1+\frac{r}{s_0}\right)^{\!2}\!
\frac{4M+q_0\!\left[r-s_0\tan^{-1}(r/s_0)\right]
}{4M\left(r^2+s_0^2\right)-q_0 s_0\!
\left[\tan^{-1}(r/s_0)\left(r^2+s_0^2\right)-r s_0 \right]
},\label{eq35}\\[5pt]
\bar b & = -\,\pi +b_R +b_D \label{eq36}
\end{align}
with
\begin{align}
b_D = \bar a \log\!\left(\frac{2\beta_p}{A_p}\right)
& = \frac{2\sqrt{2}}{\sqrt{-q_0}}
\left(1+\frac{r}{s_0}\right)^{\!2}\!
\frac{4M+q_0\!\left[r-s_0\tan^{-1}(r/s_0)\right]
}{4M\left(r^2+s_0^2\right)-q_o s_0\!
\left[\tan^{-1}(r/s_0)\left(r^2+s_0^2\right)-r s_0\right]}
\nonumber\\[5pt]
&\times
\log\!\left[
\frac{q_0 r^5\! \left(4M+(2+q_0)r-q_0 s_0\tan^{-1}(r/s_0)\right)^2
}{2\left(r^2+s_0^2\right)^2\!\left(
4M+q_0 r-q_0 s_0\tan^{-1}(r/s_0)\right)^3}
\right]\!. \label{eq37}
\end{align}

\begin{figure}[!h]
        \centerline{
        \includegraphics[scale = 0.65]{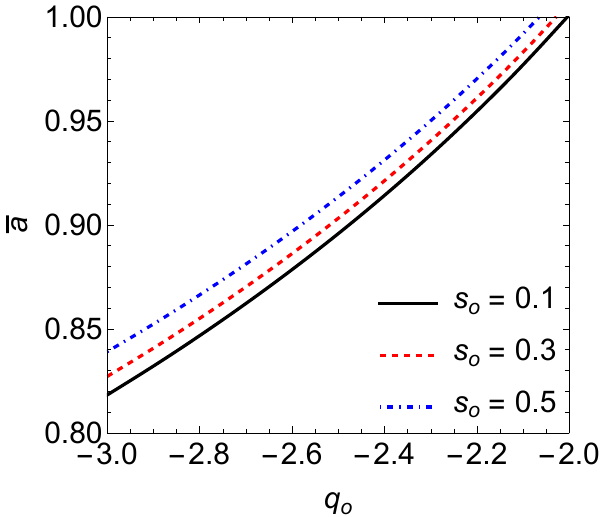}
\hspace{0.5cm}
        \includegraphics[scale = 0.65]{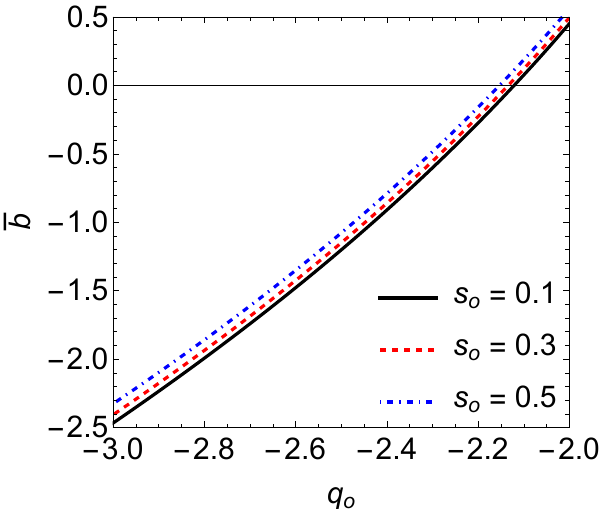}}
        \vspace{-0.2cm}
\caption{Variation of deflection coefficients $\bar{a}$ and $\bar{b}$ with 
respect to model parameter $q_0$ for different values of the model parameter
$s_0$.}
\label{fig6}
\end{figure}

In Eq.~\eqref{eq36}, $b_R$ refers to the regular part of the coefficient 
$\bar{b}$ corresponding to the integral $I_R(r_0)$ given by Eq.~\eqref{eq33}. 
To understand the behaviour of the deflection coefficients $\bar{a}$ and 
$\bar{b}$, we plot them as a function of the parameter $q_0$ for different
values of the parameter $s_0$ as shown in Fig.~\ref{fig6}. Both $\bar{a} $ 
and $\bar{b}$ show similar kinds of variation in the desired range $-3<q_0<-2$.
For this range of $q_0$ values, the deflection coefficients gradually decrease 
with decreasing values of $q_0$. From this Fig.~\ref{fig6}, one can also see 
that values of $\bar{a}$ are larger than $\bar{b}$. By analyzing the slopes of 
different curves, we can conclude that $\bar{b}$ decreases at a higher pace 
than $\bar{a}$ with decreasing values of $q_0$. It is also clearly visible 
that on increasing the value of $s_0$, values of $\bar{a}$ and $\bar{b}$ 
increase.

\begin{figure}[!h]
        \centerline{
        \includegraphics[scale = 0.6]{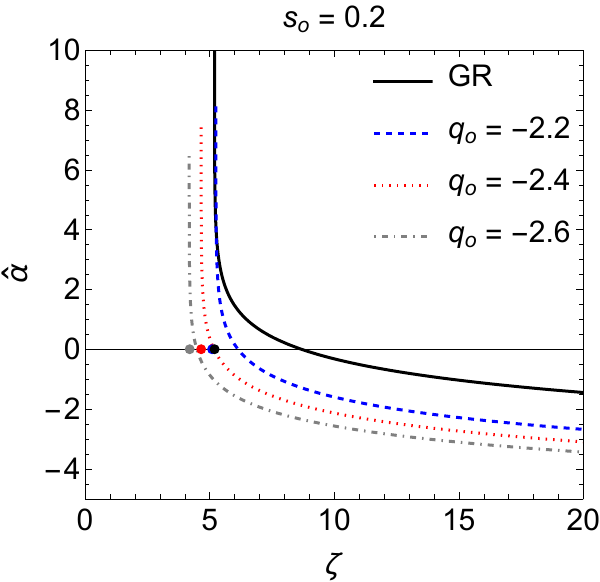}
\hspace{0.5cm}
        \includegraphics[scale = 0.6]{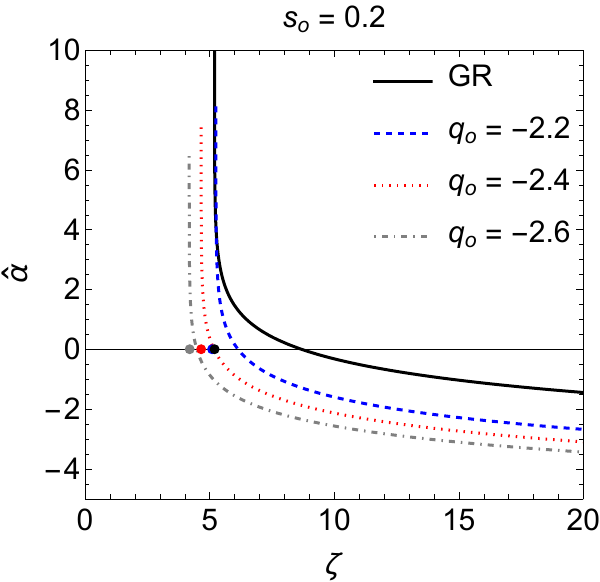}}
        \vspace{-0.2cm}
\caption{Variations of the deflection angle $\hat{\alpha}$ with the impact 
parameter $\zeta$ for different values of the model parameters 
$q_0$ and $s_0$.}
\label{fig7}
\end{figure}

Using the expressions of $\bar{a}$ and $\bar{b}$ from the respective 
Eqs.~\eqref{eq35} and \eqref{eq36}, we illustrate the variation of deflection 
angle $\hat{\alpha}$ of the strong field bending of light by the 
$f(Q,\mathcal{B})$ BH lens as a function of the impact parameter $\zeta$ for
different values of model parameters $q_0$ and $s_0$. During the analysis of 
$\hat{\alpha}$, we calculated critical impact parameter $\zeta_p$ from the 
equation $\zeta_p = r_p / \sqrt{A(r_p)}$ and have obtained the values of 
$\bar{a}$ and $\bar{b}$ at $r = r_p$. The regular contribution part $b_R$ is 
calculated for each pair of considered values of the model parameters by 
solving the regular integral numerically and added to $\bar{b}$ manually. We 
depicted our analysis of $\hat{\alpha}$ in Fig.~\ref{fig7} for different 
$q_0$ and $s_0$ values representing each plot.

From Fig.~\ref{fig7}, it is evident that we have attained the standard 
results of strong field deflection of light. From the representation of 
$\hat{\alpha}$ in the figures, we can see that the deflection angle gradually 
increases with decreasing the impact parameter, and it diverges at 
$\zeta = \zeta_p$, i.e., at photon sphere radius. In two plots of the figure, 
corresponding to two fixed values of $s_0$, we show the deflection angle 
$\hat{\alpha}$ for different $q_0$ values chosen from our desired range of 
$-3< q_o <-2$, and it clearly shows that the deflection angle, the consequence 
of the bending of light by the $f(Q,\mathcal{B})$ BH, is smaller than the 
Schwarzschild case in most of the scenarios. Moreover, we can also see that 
for a fixed value of $s_0$, the deflection angle $\hat{\alpha}$ decreases with
 decreasing the parameter $q_0$. Also from the plots of this figure, we can 
say that the deffection angle decreases by a very minute amount with increasing 
the parameter $s_0$ for a fixed $q_0$ value. Further, after a certain 
value of the impact parameter $\zeta$, depending on the values of $q_0$ and 
$s_0$, the deflection angle becomes negative, which can not be considered as 
abnormal or unphysical, as several works have already reported and 
explained the negative deflection angle \citep{71,91}. The negative 
deflection angle signifies that the photon trajectory is deflected away from 
the lensing object instead of towards it. In these cases, the negative sign 
can be interpreted as an effective repulsive gravitational effect rather than 
a breakdown of the underlying theory.

\section{Strong field observables} \label{V}
This subsection is dedicated to the estimation of the strong field observables 
under the assumption that the gravitational field of the supermassive BH 
Sgr A* at the center of our galaxy is characterized by the BHs in 
$f(Q,\mathcal{B})$ gravity theory. We analyze three most important strong 
field observables, viz., the angular position of the relativistic images 
$\vartheta$, the angular separation $s$ between the outermost relativistic 
image and the subsequent inner images which are observed to be fused together, 
and the relative magnification $r_\text{mag}$ of the outermost relativistic 
image with other relativistic images. Estimating these three key lensing 
observables in the strong field regime is very important, as their evaluation 
reveals the nature of lensing BHs. 
%
\begin{figure}[!h]
        \centerline{
        \includegraphics[scale = 0.95]{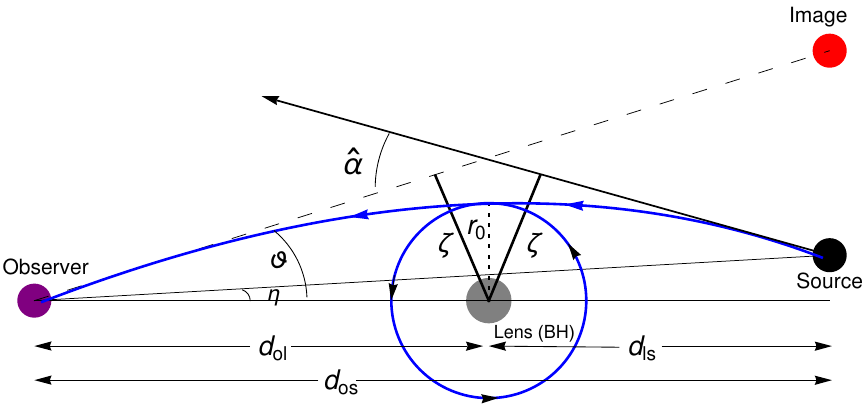}}
        \vspace{-0.2cm}
\caption{Schematic diagram of the strong field gravitational lensing by a black
hole \cite{70}.}
\label{fig8}
\end{figure}

Fig.~\ref{fig8} shows the schematic diagram of the strong field limit lensing 
by a BH. To simplify calculations, we consider that the 
source and observer are positioned at infinity from the lens. To proceed 
further, we consider a perfect alignment of the celestial bodies along the 
line of sight of the observer. This consideration is necessary to produce the 
appearance of the relativistic images as circular rings known as the Einstein 
ring. Otherwise, images on both sides of the BH appear significantly diminished 
in size. Accordingly, following Bozza's methodology, we can formulate the 
lens equation, which relates the angular positions of the source and the lens 
with the distance between the celestial bodies, as given by 
\cite{51,56,85,94,96}
\begin{equation}
\eta = \vartheta - \frac{d_\text{ls}}{d_\text{os}}\,\Delta \alpha_n, \label{eq38}
\end{equation}
where $\eta$ and $\vartheta$ are the angular positions of the source and the 
image respectively from the optic axis, $\Delta \alpha_n = \hat{\alpha} - 
2n\pi$ is the $n$th order offset deflection angle with $n$ a positive 
integer that denotes the total number of complete loops made by a photon 
around the BH. $d_\text{ls}$ and $d_\text{os}$ represent the distances of 
the source and the observer from the lens, respectively. In order to 
approximate the offset deflection from the actual deflection, we need to find 
the values of $\vartheta_n$ such that $\hat{\alpha}(\vartheta) = 2 n \pi$. 
Solving the deflection angle Eq.~\eqref{eq34} with 
$\hat{\alpha}(\vartheta) = 2 n \pi$ along with the approximation of 
$\zeta = \vartheta d_\text{ol}$, where $d_\text{ol}$ is the distance between
the observer and the lens, we find \cite{51,89,85}
\begin{equation}
 \vartheta_n^0 = \frac{\zeta_p}{d_\text{ol}}\left[1 + \exp\left(\frac{\bar{b} - 2n\pi}{\bar{a}}\right) \right].\label{eq39}
\end{equation}
%
%
%
Now, the offset angle $\Delta \alpha_n$ can be found by expanding 
$\hat{\alpha} (\vartheta)$ around $\vartheta = \vartheta_n^0$. Insertion of 
this result into the lens Eq.~\eqref{eq38} leads to the formula for image 
angle as \cite{51,85,93} 
\begin{equation}
\vartheta_n = \vartheta_n^0 + \frac{\zeta_p \exp\left(\bar{b}-2n\pi/\bar{a}\right)\left(\eta - \vartheta_n^0\right)d_\text{os}}{\bar{a}\,d_\text{ls}\,d_\text{ol}}. 
\label{eq41}
\end{equation}
\begin{figure}[!h]
        \centerline{
        \includegraphics[scale = 0.55]{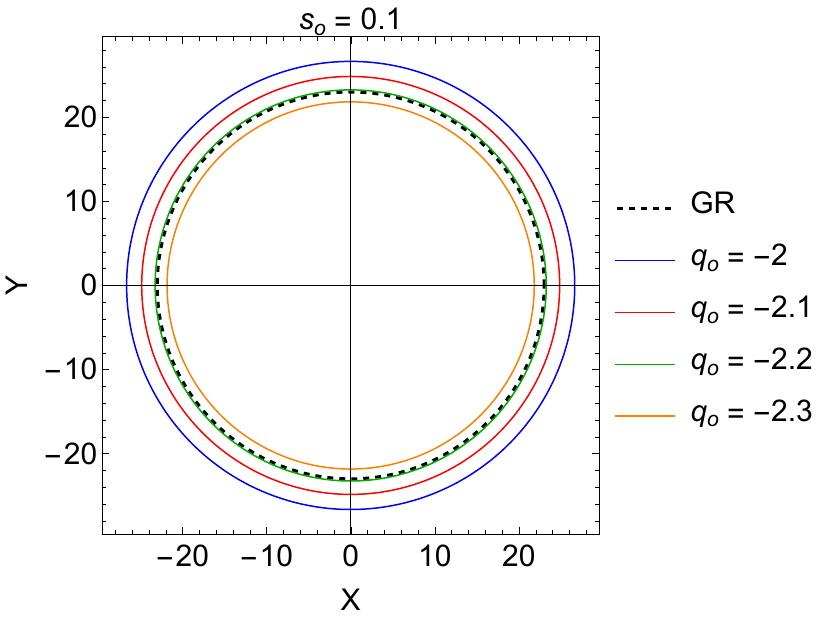}
        \hspace{0.07cm}
        \includegraphics[scale = 0.55]{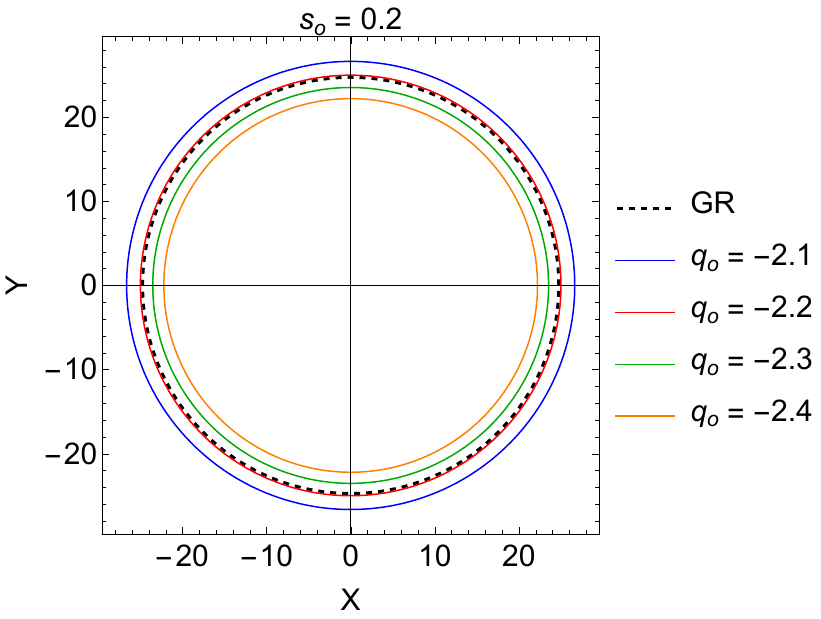}}
        \centerline{
        \hspace{0.0cm}
        \includegraphics[scale = 0.55]{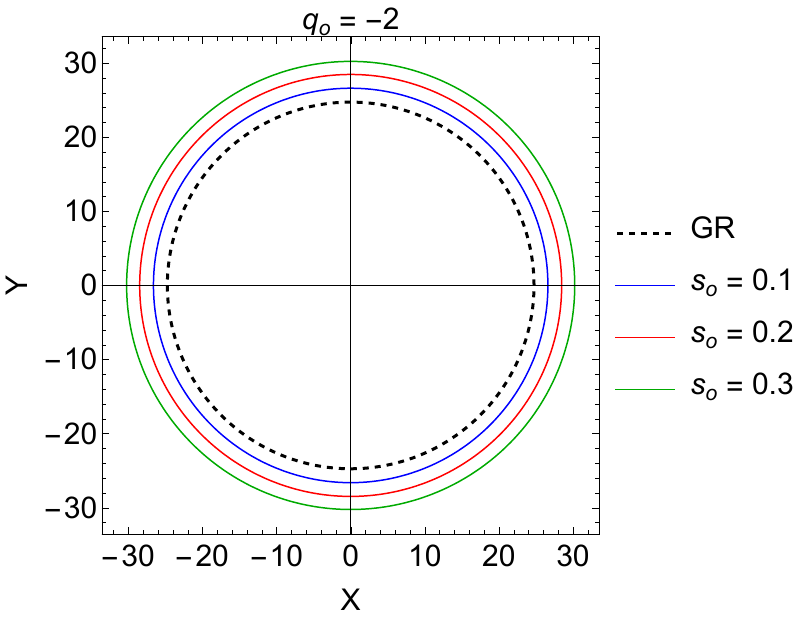}}
        \vspace{-0.2cm}
\caption{Parametric plots of the outermost Einstein ring of Sgr A* modeled
as a $f(Q,\mathcal{B})$ gravity black hole for different values of parameters
$s_0$ and $q_0$.}
\label{fig9}
\end{figure}
In Eq.~\eqref{eq41}, the correction to $\vartheta^0_n$ is much lesser than 
itself and it can be seen clearly from Eq.~\eqref{eq39} that $\vartheta_n^0$ 
falls off rapidly as $n\rightarrow \infty$, which signifies that the inner 
images are inseparable or we can say that the inner images get fused together 
and appearing as a collective image with the position approximated as 
$\vartheta_\infty = \zeta_p/d_\text{ol}$. As mentioned earlier, the perfect 
alignment of the bodies along the line of sight of the observer leads to 
maximum visibility of the images. The images thus formed are ring shaped, 
called the relativistic Einstein ring and the angular radius of the $n$th 
such ring is given by Eq.~\eqref{eq41}. By taking $n = 1$, in Eq.~\eqref{eq41},
we represent the outermost Einstein ring of the BH Sgr A* by considering it 
as an $f(Q,\mathcal{B})$ gravity BH in Fig.~\ref{fig9}. We illustrated the 
angular size of the outermost Einstein ring for different values of the 
parameters $q_0$ and $s_0$ and compared it with the size of the outermost ring 
of the Schwarzschild BH. For estimation of the angular size of Einstein 
rings, we consider the mass $M$ and distance $d_\text{ol}$ from Earth of the 
BH Sgr A* as $4.3 \times 10^{6} M_{\odot}$ and $8.35$ kpc 
($2.57\times 10^{17}$km), respectively \cite{103,104,105}. For the Sgr A* BH, 
the Einstein radius is $\vartheta^E = (25.90 \pm 1.15)$ µas as reported by the 
EHT collaboration \cite{103}. Analyzing the structure of rings for different 
parameter values, we found that the angular size of the outermost Einstein ring 
decreases with decreasing the value of $q_0$ for a fixed $s_o$ value. Moreover, 
the size of rings is seen to increase with increasing the parameter $s_0$ 
value for a fixed $q_0$ value as shown in the bottom plot of Fig.~\ref{fig9}. 

\begin{figure}[!h]
        \centerline{
        \includegraphics[scale = 0.65]{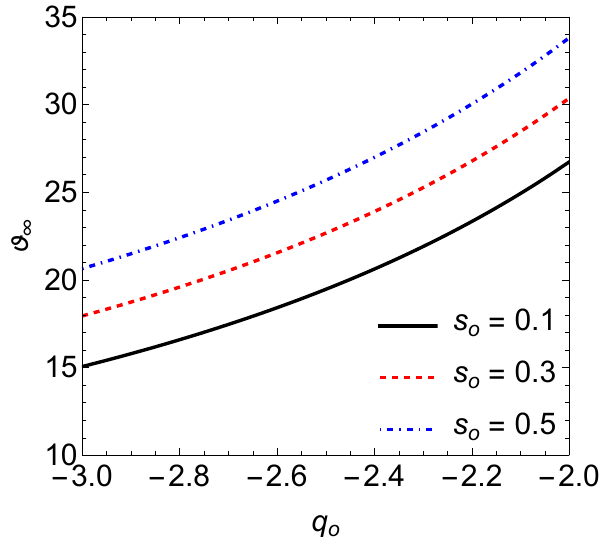}
\hspace{0.5cm}
        \includegraphics[scale = 0.68]{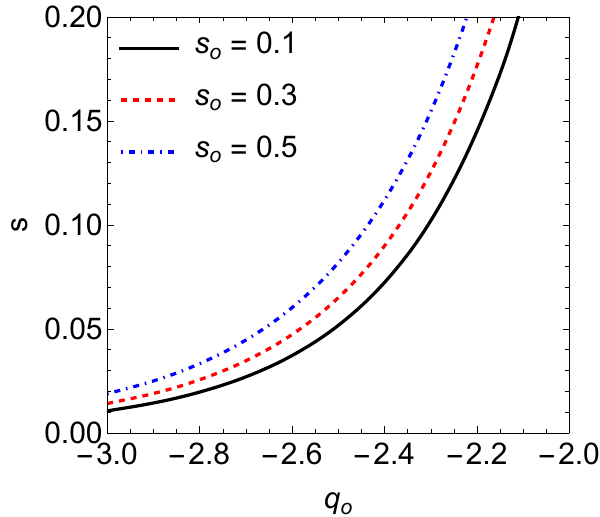}}
        \vspace{0.3cm}
\centerline{
        \includegraphics[scale = 0.65]{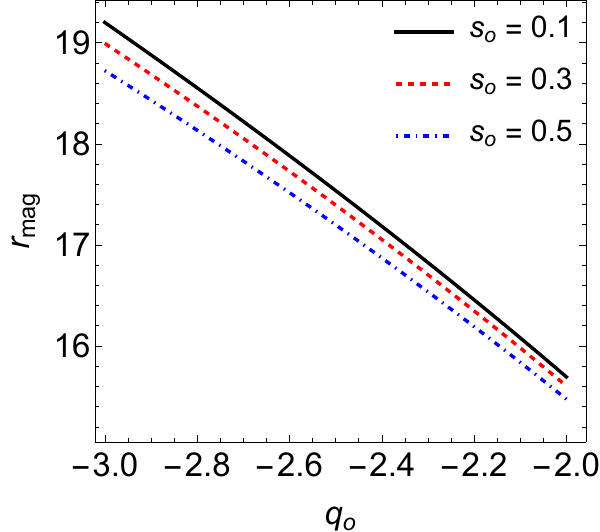}}
        \vspace{-0.2cm}
\caption{Strong field observables, viz., $\vartheta_\infty$ (left plot), $s$
(right plot) and $r_\text{mag}$ (bottom plot) as functions of parameter $q_0$
with different values of the parameter $s_0$ for the supermassive black hole
Sgr A* as an $f(Q,\mathcal{B})$ gravity black hole.}
\label{fig10}
\end{figure}
We now proceed to analyze the strong field observables as mentioned earlier. 
The angular position of the closely packed inner images $\vartheta_\infty$ 
can be obtained by taking $n\,\rightarrow\,\infty$ in Eq.~\eqref{eq41} as 
discussed already. The behaviour of this observable $\vartheta_\infty$ for
the Sgr A* as an $f(Q,\mathcal{B})$ BH is presented in the left plot of 
Fig.~\ref{fig10} as a function of the model parameter $q_0$ with different
values of the parameter $s_0$. The plot of the figure shows that 
$\vartheta_\infty$ decreases gradually with decreasing value of the parameter 
$q_0$, whereas it increases with the increasing value of the parameter $s_0$. 

The second observable $s$ is the angular separation of the first relativistic 
image from the rest of the collective inner images. As we have mentioned 
before, only the outermost image $\vartheta_1$ is distinguishable from 
the rest of the remaining inner images packed as $\vartheta_\infty$, 
Eq.~\eqref{eq41} immediately leads to the expression of $s$ as \cite{51,93,95}
\begin{equation}
s = \vartheta_1 - \vartheta_\infty = \vartheta_\infty \exp \left(\frac{\bar{b} - 2\pi}{\bar{a}}\right).\label{eq42}
\end{equation}
We analyze the behaviour of $s$ using this Eq.~\eqref{eq42} by considering the 
Sgr A* as an $f(Q,\mathcal{B})$ BH and present the result in the right 
plot of Fig.~\ref{fig10}. As in the previous case, $s$ is presented as a 
function of the model parameter $q_0$ for different values of the other 
parameter $s_0$. It is evident that the angular separation $s$ falls 
substantially with decreasing values of the parameter $q_0$. Moreover, it is 
also seen that $s$ increases with increasing $s_0$ values. Further, one 
can infer that the angular separation $s$ will eventually fall to zero 
monotonically for any fixed value of $s_0$.

The last strong field lensing observable that we include in this study is the 
relative magnification of images $r_\text{mag}$. This is defined as the ratio 
of the magnification of the outermost image to the sum of the magnifications of 
all the other inner images. The magnification of an image is an important 
source of information and is defined as the inverse of the Jacobian evaluated 
at the position of the image approximated at $\vartheta_n^o$, given by 
\cite{51,85,91}
\begin{equation}
\mu_n = \left(\frac{\eta}{\vartheta}\,\frac{\partial \eta}{\partial \vartheta} \right)^{-1}\bigg|_{\vartheta_n^0}
\approx \frac{\exp \left(\bar{b}-2n\pi/\bar{a}\right)\left[1+\exp \left(\bar{b}-2n\pi/\bar{a}\right)\right] d_\text{os}}{\bar{a}\,\eta\, d_\text{ls}\, d_\text{ol}^{\,2}}\,\zeta_p^{\,2}. \label{eq43}
\end{equation}
The equation above justifies the maximum brightness of the outermost image 
compared to the other relativistic images. Moreover, the presence of the 
$d^{\,2}_\text{ol}$ term in the denominator of Eq.~\eqref{eq43} ensures that 
the overall luminosity of all images remains relatively weak. Thus, as 
explained earlier, the relative magnification is given by \cite{51,85} 
\begin{equation}
r_\text{mag} = \frac{\mu_1}{\sum^{\infty}_{n\, =\, 2} \mu_n} = \exp \left(\frac{2\pi}{\bar{a}}\right). \label{eq44}
\end{equation}
As in the earlier cases, we estimate this key lensing observable for the 
relativistic images of the Sgr A* by considering it as an $f(Q,\mathcal{B})$ 
gravity BH. The variations of the observable $r_\text{mag}$ with the parameter 
$q_0$ for different fixed $s_0$ values are represented in the bottom plot of 
Fig.~\ref{fig10}. It is seen from the plot that the relative magnification 
$r_\text{mag}$ increases with decreasing the parameter $q_0$. Moreover, for a 
fixed value of $q_0$, the relative magnification is found to attain a 
comparatively smaller value for larger $s_0$.

\section{Black hole shadows and constraining the model parameters} \label{VI}

According to the classical picture, a BH captures all the light falling onto 
it and releases nothing. So it is obvious for an observer to expect a black 
dot or black circular area on the observer's screen, where the location of 
the BH is approximated. However, due to strong bending of light by the 
gravitational field of the BH, both the shape and size of the shadow appear 
different than our Euclidean expectations. Thus, the shadow of a BH can be 
understood as a dark silhouette of the BH against a bright background. This is 
also strongly influenced by the nature of gravitational bending of light. 
Analysis of the BH shadow hints at the size and shape of a BH as well as its
dynamic nature. Further, this analysis can be used to constrain the theories 
of gravity, taking into account the available shadow radius data of BHs from
the EHT collaboration. The study of the shadow of a BH can be made through its 
photon sphere as the shadow radius $r_\text{sh}$ and photon sphere radius 
$r_p$ are related by the relation \cite{51}:   
\begin{equation}
r_\text{sh} = \frac{r_p}{\sqrt{A(r_p)}}. \label{eq45}
\end{equation}
The basic idea behind constraining the model parameters of a BH spacetime 
solution is by comparing the theoretically calculated angular radius of the BH 
shadow with the observed angular radius of the BHs, which have recently been 
measured by the EHT collaboration, as mentioned. This method has a robust 
application for Sgr A* and is described in \cite{98}. The comparison basically 
put constrains on the theoretical data, and thus we obtain values or ranges of 
values of the model parameters where they are allowed to describe the BH 
Sgr A* \citep{99,100,101}. Like our previous discussions, in this section too 
we will use the mass $M$ and distance to lens $d_\text{ol}$ as 
$4.3 \times 10^{6} M_{\odot}$  and  $8.35$ kpc, respectively. Mass $M$ and 
distance to Sgr A*, $d_\text{ol}$ are tracked by two teams of the EHT 
collaboration, viz., `Keck' and `VLTI' \citep{103}. The second ingredient we 
require is a calibration factor connecting the size of the bright ring of 
emission with the size of the corresponding shadow, which quantifies how safe 
it is to use the size of the bright ring of emission as a proxy for the 
shadow size. The calibration factor in practice accounts for uncertainties in 
measurement, ranging from fitting/model uncertainties to theoretical 
uncertainties. In this context, the EHT teams defined a new quantity 
$\delta$, which represents the fractional deviation between the inferred 
shadow radius $r_\text{sh}$ and the shadow radius of a Schwarzschild BH 
$r_\text{sh,sc}$, given by \cite{98}
\begin{equation}
\delta = \frac{r_\text{sh}}{r_\text{sh,sc}}-1. \label{eq46}
\end{equation}
The teams, Keck and VLTI estimated the value of $\delta$ as
$$
\text{Keck: } \delta = -\,0.04^{+0.09}_{-0.10}  \;\;\; \text{and} \; \; \;
\text{VLTI: } \delta = -\,0.08^{+0.09}_{-0.09}.
$$
Being conservative, we use the average of the two estimations of $\delta$ as 
$\delta \approx -\,0.060\pm 0.065$. This value of $\delta$, under the 
Gaussianity assumption, leads to the following $1\sigma$ and $2\sigma$ 
intervals for $\delta$ \cite{98}:
$$
-\,0.125 \,\lesssim \,\delta\, \lesssim 0.005 \;\;\; \text{and} \;\;\;
-0.19 \,\lesssim\, \delta \,\lesssim 0.07. 
$$
These bounds when used in Eq.~\eqref{eq46} produces the bounds for the shadow 
radius $r_\text{sh}$ of the BH as \cite{98}
$$
4.55 \lesssim \frac{r_\text{sh}}{M} \lesssim 5.22 \,\, (1\sigma) \;\;\; 
\text{and} \;\;\; 4.21 \lesssim \frac{r_\text{sh}}{M} \lesssim 5.56 \,\, (2\sigma),
$$
where $r_\text{sh,sc} = 3\sqrt{3}M$ is used as the standard value for the 
Schwarzschild BH shadow radius.
\begin{figure}[!h]
        \centerline{
        \includegraphics[scale = 0.66]{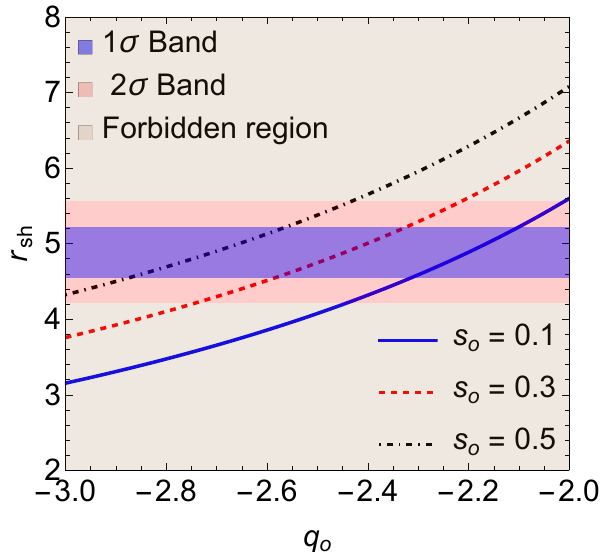}
\hspace{0.5cm}
        \includegraphics[scale = 0.65]{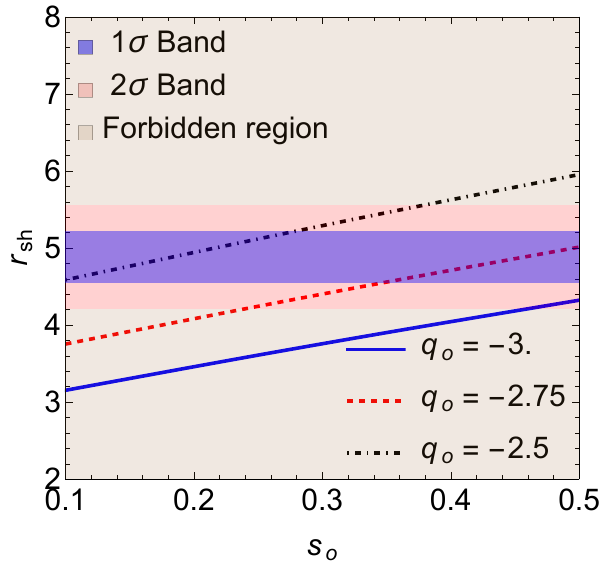}}
        \vspace{-0.2cm}
\caption{Constraining the model parameters $q_0$ and $s_0$ of the 
$f(Q,\mathcal{B})$ gravity black hole spacetime using the shadow radius data 
of Sgr A* black hole provided by EHT teams.}
\label{fig11}
\end{figure}
\begin{figure}[!h]
        \centerline{
        \includegraphics[scale = 0.48]{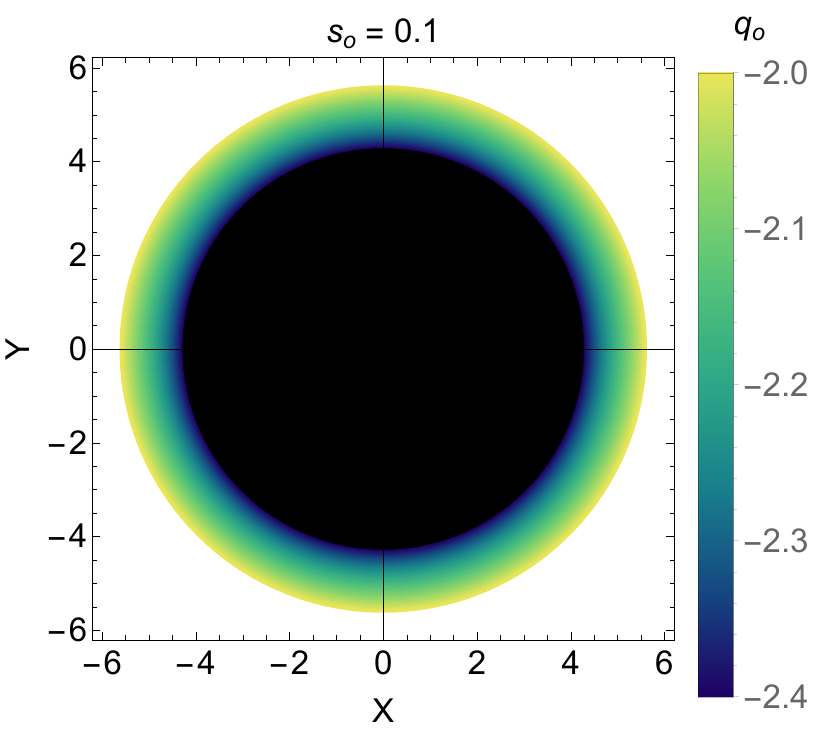}
\hspace{0.5cm}
        \includegraphics[scale = 0.48]{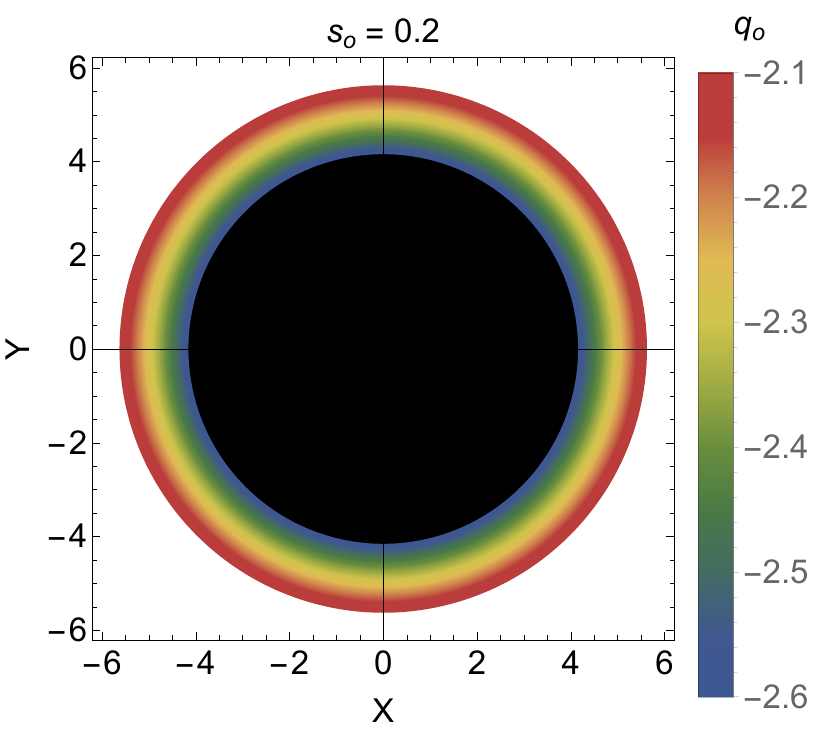}}
\hspace{0.3cm}
        \centerline{
        \includegraphics[scale = 0.48]{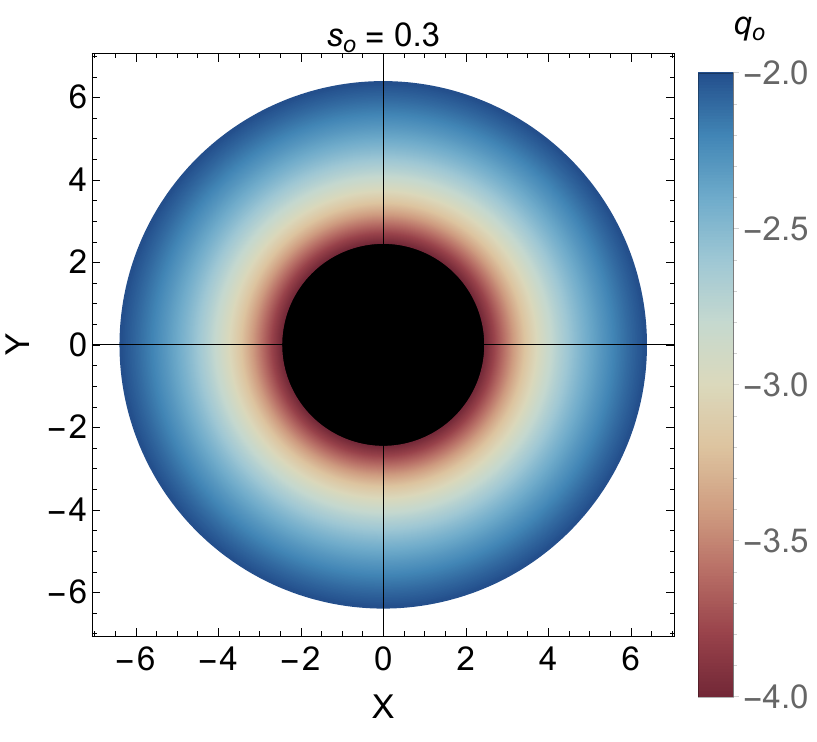}}
        \vspace{-0.5cm}
\caption{Stereoscopic mapping of the shadows of the black hole Sgr A*
by considering it as an $f(Q,\mathcal{B})$ gravity black hole.}
\label{fig12}
\end{figure}

Thus, we get two ranges of calculated angular radius of shadow for the BH 
Sgr A*. We use this data to constrain the model parameters $q_0$ and $s_0$ of 
the considered $f(Q,\mathcal{B})$ BH solution. For this, we calculated the 
shadow radius from Eq.~\eqref{eq45} as a function of the parameter $q_0$ as
well as $s_0$ for respective different fixed values of the parameters $s_0$
and $q_0$. We represent these calculated results graphically in 
Fig.~\ref{fig11}. The inclusion of the $1\sigma $ and $2\sigma $ bands in the 
shadow radius plots limits the allowed values of $q_0$ and $s_0$. The left 
plot of Fig.~\ref{fig11} 
shows variations of shadow radius as a function of $q_0$ for different 
fixed values of $s_0$, and the right plot of it shows the variation of 
shadow radius as a function of $s_0$ for different fixed values of $q_0$. We 
represent the $1\sigma$ band by the red shaded region and the $2\sigma$ band 
by the blue shaded region, while the brown shaded areas represent the 
forbidden region for the shadow radius. 
It is also visible from Fig.~\ref{fig11} that the shadow radius increases with 
increasing the values of the parameters $q_0$ and $s_0$. Moreover, we can 
clearly observe that the allowed range of $q_0$ enlarges slightly with 
increasing $s_0$ values. Further, the figure shows that the shadow radius for 
such values of $s_0$, the range $-3 \lesssim q_0 \lesssim -2$ includes most 
of the allowed values of $q_0$. This is why we have selected this range of 
$q_0$ as the desired range in our studies in previous sections. 

For a static and spherically symmetric spacetime solution, the shadow only 
differs in size and not in shape than their imaginary Euclidean 
representations. To determine as well as visualize the shadow's apparent size, 
we use the stereographic projection of the BH's shadow from the plane of the 
black hole to the observer's observation plane using coordinates $(X,Y)$, and 
these are defined as \cite{102} 
\begin{center}$
X = \underset{r_O\rightarrow \infty}{ \lim} \left(-r^2_o \sin \vartheta_o \frac{d\phi}{dr}\!\mid_{r_o}\right) \;\;\; \text{and} \; \;\;
Y = \underset{r_o\rightarrow \infty} {\lim} \left(  r^2_o \frac{d\vartheta}{dr}\mid_{(r_o,\vartheta_o)}\right)\!, 
$\end{center}
where $r_o$ and $\vartheta_o$ are the radial position and angular position of 
the observer with respect to the BH plane. Fig.~\ref{fig12} shows the 
stereographic projection of the shadows of the $f(Q,\mathcal{B})$ BH as a 
function of the model parameter $q_0$ for different fixed values of $s_0$. 
The left and right plots of Fig.~\ref{fig12} show the variation of shadow 
size of the considered $f(Q,\mathcal{B})$ BH for the allowed range of $q_0$ 
keeping $s_0$ fixed at $0.1$ and $0.2$, respectively. Close observation of 
the stereotypic projection reveals that the shadow size decreases with 
increasing negative values of $q_0$. Moreover, we also provide the 
stereoscopic mapping of the angular radius of shadow of the BH for a broader 
range of $q_0$ with $s_0$ fixed at $0.3$, represented by the bottom plot of 
Fig.~\ref{fig12}. All these stereoscopic mappings show that the shadow of the
BH decreases with the increasing values of $s_0$ also. 

\section{Summery and Conclusion} \label{VII}
Gravitational lensing is a versatile technique for exploring the properties of 
spacetime and for testing the predictions of different theories of gravity. In 
this work, we have studied the gravitational lensing of light by an 
$f(Q,\mathcal{B})$ gravity BH. First, we have discussed 
the horizon structure of the BH solution by analyzing its behavior as a 
function of the model parameters $q_0$ and $s_0$. Then, we moved to the the 
estimation of strong field deflection angle. For this, we have used the 
standard method developed by V.~Booza, which is applicable to all 
spacetime in any gravity theory. Following this method, we first obtained the 
deflection coefficients $\bar{a}$ and $\bar{b}$, and next using them we 
formulate the equation for deriving the deflection angle. Plotting the 
deflection angle against impact parameter showed that we have obtained the 
standard result, i.e, the deflection angle diverges as the impact parameter 
approaches its critical value. This shows that a photon with critical impact 
parameter have infinite deflection angle, i.e, the photon will encircle the 
BH for the rest of eternity. Deflection angle in our model is dependent upon 
the model parameters $q_0$ and $s_0$. It is found to be decreasing for 
decreasing $q_0$ values and goes negative for large values of impact parameter, 
highlighting repulsive interactions by the gravitational field of the BHs. No 
abrupt change of deflection angle from the Schwarzschild case is recorded and 
only a small variations from this case is found.

We have also computed the numerical values of the 
lensing observables: the angular position $\vartheta_\infty$ of relativistic 
images, angular separation $s$, and relative magnification $r_\text{mag}$, by 
considering the supermassive BH SgrA* as an $f(Q,\mathcal{B})$ gravity BH. We 
portrayed the outermost rings of the BH Sgr A* with the help of its 
measured mass $M$ and its distance from Earth, $d_\text{ol}$. All of the 
lensing observables showed mostly the same type of behavioural decrease with 
increasing the negative value of the parameter $q_0$ for a fixed $s_0$, except 
the observable $r_\text{mag}$, which showed opposite behaviour. 

The last part of our work is dedicated to the study of the shadow of the
considered BH solution. For this, we have used the data provided by the EHT 
collaboration on BH Sgr A* at the center of our galaxy. The two teams of EHT 
collaboration, viz., `keck' and 
`VLCI' have independently obtained the two ranges of angular radius of the 
shadow of the BH. From the average results of these two teams, $1\sigma$ and
$2\sigma$ bounds on the shadow radius of BHs can be obtained. Using these 
bounds on the shadow radius, we constrained the model parameters $q_0$ and 
$s_0$. The corresponding representation of theoretically calculated data with 
EHT observed data reveals that the $f(Q,\mathcal{B})$ gravity BH is valid for 
every considered values of $s_0$. But for each $s_0$, there exists a 
particular range of 
$q_0$, inside which it can mimic the BH Sgr A*. The range of allowed $q_0$ is 
found to be gradually increasing with increasing $s_0$. With this in hand, we 
make the stereoscopic projection of the BH Sgr A* by considering it as an 
$f(Q,\mathcal{B})$ gravity BH. The shadow size of this BH decreases
with increasing negative values of $q_0$ as well as increasing values 
of $s_0$.

This study seeks to enhance our understanding of the 
$f(Q,\mathcal{B})$ gravity theory and its role in describing gravitational 
phenomena beyond GR. Through the analysis of strong gravitational lensing
and shadow, we aim to explore the observational consequences of this theory 
and contribute to the growing body of research on ATGs.


\begin{thebibliography}{100}

\bibitem{1} C.M. Will, \emph{The confrontation between General Relativity and Expermient}, \href{https://doi.org/10.12942/lrr-2014-4}{Living Reviews in Relativity \textbf{17}, 4 (2014)}.

\bibitem{2} B.P. Abbott et al. ,\emph{Observation of Gravitational Waves from a Binary BlackHole Merger}, \href{ https://doi.org/10.1103/PhysRevLett.116.061102}{Phys. Rev. Lett.\textbf{116}, 061102 (2016)}.

\bibitem{3} B.P. Abbott et al. \emph{Observation of Gravitational Waves from a 22-Solar-Mass Binary Black Hole Coalescence}, \href{https://doi.org/10.1103/PhysRevLett.116.241103}{Phys. Rev. Lett. \textbf{116}, 241103 (2016)}.

\bibitem{4} B.P. Abbott et al., \emph{Observation of Gravitational Waves from a Binary Neutron Star Inspiral}, \href{ https://doi.org/10.1103/PhysRevLett.119.161101}{Phys. Rev. Lett.\textbf{119}, 161101 (2017)}.

\bibitem{5} B.P. Abbott et al., \emph{Observation of a Binary-Black-Hole Coalescence with Asymmetric Masses},\href{ https://doi.org/10.1103/PhysRevD.102.043015}{Phys. Rev. D \textbf{102}, 043015 (2020)}.

\bibitem{6} R. Abbott et al., \emph{Observation of Gravitational Waves from Two Neutron Star–Black Hole Coalescences}, \href{https://doi.org/10.3847/2041-8213/ac082e}{APJL\textbf{915}, L5 (2021)}.

\bibitem{7} R. Abbott et al., \emph{Search for continuous gravitational wave emission from the MilkyWay center in O3 LIGO-Virgo data}, \href{https://doi.org/10.1103/PhysRevD.106.042003}{Phys. Rev. D \textbf{106} 042003 (2022)}.

\bibitem{8}Event Horizon Telescope Collaboration et al.,\emph{First M87 Event Horizon Telescope Results. I.The Shadow of the Supermassive Black Hole}, \href{https://doi.org/10.3847/2041-8213/ab0ec7}{APJL \textbf{875}, L1 (2019)}.

\bibitem{9} Event Horizon Telescope Collaboration et al., \emph{First M87 Event Horizon        Telescope Results. II. Array and Instrumentation}, \href{https://doi.org/10.3847/2041-8213/ab0c96}{APJL\textbf{875}, L2 (2019)}.

\bibitem{10} Event Horizon Telescope Collaboration et al., \emph{First M87 Event Horizon Telescope Results. III. Data Processing and Calibration}, \href{https://doi.org/10.3847/2041-8213/ab0c57}{APJL \textbf{875}, L3 (2019)}.

\bibitem{11} EventHorizonTelescopeCollaborationetal., \emph{First M87 Event Horizon Telescope Results. IV. Imaging the Central Supermassive Black Hole}, \href{https://doi.org/10.3847/2041-8213/ab0e85}{APJL \textbf{875}, L4 (2019)}.

\bibitem{12} EventHorizonTelescopeCollaborationetal., \emph{First M87 Event Horizon Telescope Results. V. Physical Origin of the Asymmetric Ring}, \href{https://doi.org/10.3847/2041-8213/ab0f43}{APJL \textbf{875}, L5 (2019)}.

\bibitem{13} Event Horizon Telescope Collaboration et al., \emph{First M87 Event Horizon Telescope Results. VI. The Shadow and Mass of the Central BlackHole}, \href{https://doi.org/10.3847/2041-8213/ab1141}{APJL \textbf{875}, L6 (2019)}.

\bibitem{14}K.S. Stelle, \emph{Renormalization of higher derivative quantum gravity}, \href{https://doi.org/10.1103/PhysRevD.16.953}{Phys. Rev. D\textbf{16}, 953 (1977)}.

\bibitem{15} N. A. Bahcall, J. P. Ostriker, S. Perlmutter and P. J. Steinhardt, \emph{The cosmic triangle: Revealing the state of the universe}, \href{https://doi.org/10.1126/science.284.5419.1481}{Science \textbf{284}, 5419 (1999)}.

\bibitem{16}Riess et al., \emph{Observational Evidence from Supernovae for an Accelerating Universe and a Cosmological Constant}, \href{https://doi.org/10.1086/300499}{The Astrono. jour \textbf{116}, 1009 (1998)}.

\bibitem{16a}V.C. Rubin,W.K. Ford and N. Thonnard, \emph{Rotational Properties of 21 Sc Galaxies with a Large Range of Luminosities and Radii}, \href{https://doi.org/10.1086/300499}{The Astrop. jour. \textbf{238}, 478487 (1980)}.

\bibitem{16b}V.C. Rubin, N. Thonnard and W.K. Ford, \emph{Extended Rotation Curves of High-Luminosity Spiral Galaxies. IV. Systematic Dynamical Properties, SA Through SC}, \href{https://doi.org/10.1086/300499}{The Astrop. jour. lett. \textbf{225}, 10711 (1978)}.

\bibitem{16c}Reiss et.al.,\emph{Observational Evidence from Supernovae for an Accelerating Universe and a Cosmological Constant},\href{https://doi.org/10.1086/300499}{Astron. Jour.\textbf{116}, 10091038 (1998)}.

\bibitem{16d} E.J. Copeland, M. Sami and S.Tsujikawa ,\emph{Dynamics of dark energy},\href{https://doi.org/10.1142/S021827180600942X}{IJMPD \textbf{15}, 17531936 (2006)}.

\bibitem{16e}G. Bertone and T.M.P. Tait ,\emph{A new era in the search of Dark Matter},\href{https://doi.org/10.1038/s41586-018-0542-z}{Nature \textbf{562}, 5156 (2018)}.

\bibitem{17} Carpil et al., \emph{Quantum Gravity: A Brief History of Ideas and Some Prospects}, \href{https://doi.org/10.1142/S0218271815300281}{IJMPD \textbf{24}, 1530028(2015)}.

\bibitem{18} A. A. Starobinsky, \emph{The Perturbation Spectrum Evolving from a Nonsingular Initially De-Sitter Cosmology and the Microwave Background Anisotropy}, \href{https://inspirehep.net/literature/199078}{Sov. Astron. Lett.\textbf{9}, 302 (1983)}.


\bibitem{19} W. Hu, I. Sawicki, \emph{ Models of f(R) Cosmic Acceleration that Evade Solar-System Tests}, \href{https://doi.org/10.1103/PhysRevD.76.064004}{Phys. Rev. D \textbf{76}, 064004 (2007)}.

\bibitem{20} D. J. Gogoi and U. D. Goswami, \emph{A new f(R) gravity model and properties of gravitational waves in it}, \href{https://doi.org/10.1140/epjc/s10052-020-08684-3}{EPJC \textbf{80}, 1101(2020)}.

\bibitem{21} P. Rastall, \emph{ Generalization of the Einstein Theory}, \href{ https://doi.org/10.1103/PhysRevD.6.3357}{Phys. Rev. D\textbf{6}, 3357 (1972)}.

\bibitem{22} F. S. N. Lobo and T. Harko, \emph{ Extended $f(R,L_M)$ theories of gravity}, \href{https://doi.org/10.1142/9789814623995_0110}{The Thirteenth Marcel Grossmann Meeting pp. 1164-1166 (2015)}.

\bibitem{23}C. Masud,Kluson, J.,,O. Markku, T. Anca, \emph{Can TeVeS be a viable theory of gravity?}, \href{https://doi.org/10.1016/j.physletb.2014.06.036}{ Phy. Lett. B \textbf{735}, 322326 (2014)}.


\bibitem{24}F. Bourliot, P.G. Ferreira, D.F. Mota and C. Skordis,\emph{Cosmological behavior of Bekenstein's modified theory of gravity}, \href{https://link.aps.org/doi/10.1103/PhysRevD.75.063508}{Phys. Rev. D \textbf{75}, 063508 (2007)}.

\bibitem{25} H. Shabani, A. De, T. H. Loo and E. N. Saridakis,\emph{ Cosmology of f(Q) gravity in non-flat Universe}, \href{https://doi.org/10.1140/epjc/s10052-024-12582-3}{EPJC \textbf{84}, 285 (2024)}.

\bibitem{26} P. Sarmah, A. De and U. D. Goswami, \emph{Anisotropic LRS-BI Universe with f(Q) gravity theory}, \href{https://doi.org/10.1016/j.dark.2023.101209}{Phys. Dark Universe    \textbf{40}, 101209 (2023)}.

\bibitem{26a}D. Zhao , \emph{Covariant formulation of f(Q) theory}, \href{https://doi.org/10.1140/epjc/s10052-022-10266-4}{    Eur. Phys. J. C \textbf{82}, 303 (2022)}.

\bibitem{26b}
W. Wang, H. Chen and T. Katsuragawa,
\emph{Static and spherically symmetric solutions in $f(Q)$ gravity},
\href{https://doi.org/10.1103/PhysRevD.105.024060}
{Phys. Rev. D \textbf{105}, 024060 (2022)}.

\bibitem{27}G.R. Bengochea and R. Ferraro, \emph{Dark torsion as the cosmic speed-up}, \href{https://link.aps.org/doi/10.1103/PhysRevD.79.124019}{Phys. Rev. D\textbf{79}, 124019 (2009)}.

\bibitem{28} A. Einstein, \emph{Riemann-Geometrie mit Aufrechterhaltung des Begriffes des Fernparallelismus}, \href{https://doi.org/10.1002/3527608958.ch36}{ Abhandlungen der Königlich Preussischen Akademie der Wissenschaften   \textbf{27}, 123130 (1928)}.

\bibitem{29}M. Adak, \emph{The Symmetric teleparallel gravity}, \href{https://doi.org/10.48550/arXiv.gr-qc/0611077}{Turk. J. Phys.\textbf{30}, 379390 (2006)}.

\bibitem{30}J.B. Jiménez, L. Heisenberg and T. Koivisto, \emph{Coincident general relativity}, \href{https://link.aps.org/doi/10.1103/PhysRevD.98.044048}{Phys. Rev. D \textbf{98}, 044048 (2018) }.

\bibitem{31}  L. Heisenberg, \emph{A systematic approach to generalisations of General Relativity and their cosmological implications}, \href{https://doi.org/10.1016/j.physrep.2018.11.006}{Phys. Rept.\textbf{796}, 1113 (2019)}.

\bibitem{32} J.B. Jimenez, L. Heisenberg, D. Iosifidis, A.J.Cano and T. S. Koivisto, \emph{General teleparallel quadratic gravity}, \href{https://doi.org/10.1016/j.physletb.2020.135422}{phys. Lett. B \textbf{805}, 135422 (2020)}.

\bibitem{33} J. Lu,X. Zhao and G. Chee, \emph{ Cosmology in symmetric teleparallel gravity and its dynamical system}, \href{https://doi.org/10.1140/epjc/s10052-019-7038-3}{EPJC \textbf{79}, 530 (2019)}.

\bibitem{34} A. Lymperis, \emph{Late-time cosmology with phantom dark-energy in f(Q) gravity}, \href{https://doi.org/10.1088/1475-7516/2022/11/018}{JCAP \textbf{11}, 018 (2022)}.

\bibitem{35} B. C. Paul, A. Chanda, A. Beesham, and S. D. Maharaj, \emph{Late time cosmology in-gravity with interacting fluids}, \href{https://doi.org/10.1088/1361-6382/ac4b97}{Class. Quant. Grav. \textbf{39}, 065006 (2022)}.

\bibitem{36} S. A. Narawade, S. P. Singh and B. Mishra, \emph{Accelerating cosmological models in f(Q) gravity and the phase space analysis}, \href{https://doi.org/10.1016/j.dark.2023.101282}{Phys. Dark Univ. \textbf{42}, 101282 (2023)}.

\bibitem{37} N. Dimakis, M. Roumeliotis, A. Paliathanasis and T. Christodoulakis, \emph{Anisotropic Solutions in Symmetric
Teleparallel f (Q)-theory: Kantowski-Sachs and Bianchi III LRS Cosmologies}, \href{https://doi.org/10.1140/epjc/s10052-023-11964-3}{EPJC    \textbf{83}, 74 (2023)}.

\bibitem{38} O. Sokoliuk, S. Arora, S. Mraharaj, A. Baransky and P. K. Sahoo,    \emph{On the impact of f(Q) gravity on the large scale structure}, \href{ https://doi.org/10.1093/mnras/stad968}{Mon. Not. Roy. Astron. Soc   \textbf{522}, 252267 (2023)}.

\bibitem{39}  M. Milgrom, \emph{Noncovariance at low accelerations as a route to MOND}, \href{ https://doi.org/10.1103/PhysRevD.100.084039}{Phys. Rev. D \textbf{100}, 084039 (2019)}.

\bibitem{40} F. D’Ambrosio, M. Garg and L. Heisenberg, \emph{Non-linear extension of non-metricity scalar for MOND}, \href{https://doi.org/10.1016/j.physletb.2020.135970}{Phys. Lett. B \textbf{811}, 135970 (2020)}.

\bibitem{41} F. Bajardi, D. Vernieri, and S. Capozziello, \emph{Bouncing Cosmology in f(Q) Symmetric Teleparallel Gravity}, \href{https://doi.org/10.1140/epjp/s13360-020-00918-3}{EPJ Plus \textbf{135}, 912 (2020)}.

\bibitem{42} A. S. Agrawal, L. Pati, S. K. Tripathy, and B. Mishra,    \emph{Matter bounce scenario and the dynamical aspects in f(Q,T) gravity}, \href{https://doi.org/10.1016/j.dark.2021.100863}{Phys. Dark Univ.\textbf{33}, 100863 (2021)}.

\bibitem{43}N. Dimakis, A. Paliathanasis, and T. Christodoulakis,    \emph{Quantum cosmology in f(Q) theory}, \href{https://doi.org/10.1088/1361-6382/ac2b09}{Class. Quant. Grav   \textbf{38}, 225003 (2021)}.

\bibitem{44} S. Capozziello, V. De Falco and C. Ferrara, \emph{The role of the boundary term in f(Q,B) symmetric teleparallel gravity}, \href{https://doi.org/10.1140/epjc/s10052-023-12072-y}{EPJC \textbf{83}, 915 (2023)}.

\bibitem{45}  A. Paliathanasis,   \emph{Symmetric teleparallel cosmology with boundary corrections}, \href{https://doi.org/10.1016/j.dark.2023.101388}{Phys. Dark Univ. \textbf{43}, 101388 (2024)}.

\bibitem{46} A. Pradhan, A. Dixit, M. Zeyauddin and S. Krishnannair, \emph{A flat FLRW dark energy model in f(Q,C)-gravity theory with observational constraints}, \href{https://doi.org/10.1142/S0219887824501676}{IJGMMP \textbf{21}, 2450167 (2024)}.

\bibitem{47}  D. C. Maurya, \emph{Quintessence behaviour dark energy models in f(Q,B)-gravity theory with observational constraints}, \href{https://doi.org/10.1016/j.ascom.2024.100798}{Astro. and Comp.\textbf{46}, 100798 (2024)}.

\bibitem{48} A. Einstein, \emph{ The Foundation of the General Theory of Relativity}, \href{https://doi.org/10.1002/andp.19163540702}{Ann. Phys. (N.Y.) \textbf{49}, 769 (1916); \textbf{14}, 517 (2005)}.

\bibitem{49}  F. W. Dyson, et al.,   \emph{ A determination of the deflection of light by the sun’s gravitational field, from observations made at the total eclipse of May 29, 1919}, \href{https://doi.org/10.1098/rsta.1920.0009}{Phil. Trans. R. Soc. A \textbf{220}, 291 (1920)}.

\bibitem{50} C. G. Darwin, \emph{The Gravity Field of a Particle}, \href{https://doi.org/10.1098/rspa.1959.0015}{Proc. A \textbf{249}, 1257 (1959)}.

\bibitem{50a} N. Parbin, D.J. Gogoi, and U.D. Goswami, \emph{Weak gravitational lensing and shadow cast by rotating black holes in axionic Chern-Simons theory}, \href{
https://doi.org/10.48550/arXiv.2305.09157}{Phys. of dark Uni.\textbf{41}, 101265 (2023)}.


\bibitem{51} V. Bozza, \emph{ Gravitational lensing in the strong field limit}, \href{https://link.aps.org/doi/10.1103/PhysRevD.66.103001}{Phys. Rev. D \textbf{66}, 103001 (2002)}.

\bibitem{52} V. Bozza, \emph{ Quasiequatorial Gravitational Lensing by Spinning Black Holes in the Strong Field Limit}, \href{https://link.aps.org/doi/10.1103/PhysRevD.67.103006}{Phys. Rev. D \textbf{67}, 103006 (2003)}.

\bibitem{53} M. Bartelmann    \emph{ Gravitational Lensing}, \href{https://doi.org/10.1088/0264-9381/27/23/233001}{Class. Quan. Grav    \textbf{27}, 233001 (2010)}.

\bibitem{54}  J. Wambsganss,   \emph{Gravitational Lensing in Astronomy}, \href{https://doi.org/10.12942/lrr-1998-12}{Living Rev. Relativ \textbf{1}, 12 (1998)}.

\bibitem{55} K. S. Virbhadra, D. Narasimha and S. M. Chitre,    \emph{ Role of the scalar field in gravitational lensing}, \href{arXiv:astro-ph/9801174}{Astro phy. As \textbf{337}, 1 (1998)}.

\bibitem{56} K. S. Virbhadra and G. F. R. Ellis, \emph{Schwarzschild black hole lensing}, \href{ https://doi.org/10.1103/PhysRevD.62.084003}{Phys. Rev. D \textbf{62}, 084003 (2000) }.

\bibitem{57} K. S. Virbhadra,    \emph{Relativistic Images of Schwarzschild Black Hole Lensing}, \href{ https://doi.org/10.1103/PhysRevD.79.083004}{ Phys. Rev. D   \textbf{79}, 083004 (2009)}.

\bibitem{58}  K. S. Virbhadra,    \emph{ Distortions of Images of Schwarzschild Lensing}, \href{https://doi.org/10.1103/PhysRevD.106.064038}{Phys. Rev. D \textbf{106}, 064038 (2002)}.

\bibitem{59}  F. Atamurotov and S. G. Ghosh,   \emph{ Gravitational Weak Lensing by a Naked Singularity in Plasma}, \href{https://doi.org/10.1140/epjp/s13360-022-02885-3}{EPJ Plus \textbf{137}, 662 (2022)}.

\bibitem{60}  Y. Chen et al.,   \emph{ Gravitational Lensing by Born-Infeld Naked Singularities}, \href{https://doi.org/10.1103/PhysRevD.109.084014}{Phys. Rev. D \textbf{109}, 084014 (2024)}.

\bibitem{61} V. Bozza and L. Mancini,    \emph{ Time Delay in Black Hole Gravitational Lensing as a Distance Estimator}, \href{https://doi.org/10.1023/B:GERG.0000010486.58026.4f}{Gen. Relativ. Gravit \textbf{36}, 435 (2004)}.

\bibitem{62} T. Hsieh, D. S. Lee, and C. Y. Lin,    \emph{Gravitational Time Delay Effects by Kerr and Kerr-Newman Black Holes in Strong Field Limits}, \href{https://doi.org/10.1103/PhysRevD.104.104013}{Phys. Rev. D \textbf{104}, 104013 (2021)}.

\bibitem{62a} L. Zhao, M. Tang and Z. Xu  ,\emph{The lensing effect of quantum-corrected black hole and parameter constraints from EHT observations},\href{https://doi.org/10.1140/epjc/s10052-024-13342-z}{EPJC \textbf{84}, 971 (2024)}.

\bibitem{62b}M.Y. Guo et.al.,\emph{Strong gravitational lensing effects around rotating regular black holes},\href{https://doi.org/10.1016/j.physletb.2024.139211}{Phys. Lett. B \textbf{860}, 139221 (2025)}.

\bibitem{63} Chen-Kai Qiao and Mi Zhou   \emph{Gravitational lensing of Schwarzschild and charged black holes immersed in perfect fluid dark matter halo}, \href{https://doi.org/10.1088/1475-7516/2023/12/005}{JCAP \textbf{2023}, 005 (2023)}.

\bibitem{64}G.S.B.Kogan and O. Y. Tsupko, \emph{Gravitational lensing in plasmic medium}, \href{https://doi.org/10.1134/S1063780X15070016}{Plas. Phys. Rep. \textbf{41}, (2015)}.

\bibitem{65}G.S.B.Kogan and O. Y. Tsupko, \emph{Gravitational lensing in plasma: Relativistic images at homogeneous plasma}, \href{https://link.aps.org/doi/10.1103/PhysRevD.87.124009}{Phys. Rev. D \textbf{87}, 124009, (2013)}.

\bibitem{66}G.S.B.Kogan and O. Y. Tsupko, \emph{Gravitational lensing in a non-uniform plasma}, \href{https://doi.org/10.1111/j.1365-2966.2010.16290.x}{Mon. Not. of the R.Astro. Soc. \textbf{404}, (2010)}.

\bibitem{67} R. Shaikh et al.,   \emph{Strong Gravitational Lensing by Wormholes}, \href{https://doi.org/10.1088/1475-7516/2019/07/028}{JCAP \textbf{2019}, 028 (2019)}.

\bibitem{68} S. Kumar, A. Uniyal, and S. Chakrabarti,    \emph{Shadow and Weak Gravitational Lensing of Rotating Traversable Wormhole in Nonhomogeneous Plasma Spacetime}, \href{ https://doi.org/10.1103/PhysRevD.109.104012}{Phys. Rev. D \textbf{109}, 104012 (2024)}.

\bibitem{69} N. Godani and G. C. Samanta, \emph{ Gravitational Lensing for Wormhole with Scalar Field in f(R) Gravity}, \href{https://doi.org/10.1142/S0219887823500755}{IJGMMP \textbf{20}, 2350075 (2023)}.

\bibitem{70}G. Mohan,R. Karmakar,R.J.Borah and U.D. Goswami, \emph{Strong lensing effect and quasinormal modes of oscillations of black holes in f(R,T) gravity theory}, \href{https://www.sciencedirect.com/science/article/pii/S2212686425002006}{Phy Dark Uni. \textbf{49}, 102007 (2025)}.

\bibitem{71}G. Mohan,N. Parbin and U.D. Goswami, \emph{. Investigating the effects of gravitational lensing by Hu-Sawicki f(R) gravity black holes}, \href{https://doi.org/10.1140/epjc/s10052-025-14143-8}{EPJC \textbf{85}, 413 (2025)}.

\bibitem{72}R. Kumar and S.G.Ghosh, \emph{Black Hole Parameter Estimation from Its Shadow}, \href{https://doi.org/10.3847/1538-4357/ab77b0}{Astro. Jour.   \textbf{892}, 78 (2020)}.

\bibitem{73} S. Capozziello, V. De Falco and C. Ferrara,    \emph{The role of the boundary term in f(Q,B) symmetric teleparallel gravity}, \href{https://doi.org/10.1140/epjc/s10052-023-12072-y}{EPJC \textbf{83}, 915 (2023)}.

\bibitem{74}  A. De, T. H. Loo and E. N. Saridakis,   \emph{Non-metricity with bounday terms: f(Q,C) gravity and cosmology}, \href{https://doi.org/10.1088/1475-7516/2024/03/050}{JCAP \textbf{2024}, 050 (2024)}.

\bibitem{75} F. D’Ambrosio, S. D. B. Fell, L. Heisenberg and S. Kuhn, \emph{Black holes in f(Q) gravity}, \href{https://doi.org/10.1103/PhysRevD.105.024042}{Phys. Rev. D \textbf{105}, 024042 (2022)}.

\bibitem{76}D. Zhao, \emph{Covariant formulation of f(Q) theory}, \href{https://doi.org/10.1140/epjc/s10052-022-10266-4}{EPJC \textbf{82}, 303 (2022)}.

\bibitem{77} R. H. Lin and X. H. Zhai,  \emph{Spherically symmetric configuration in f(Q) gravity}, \href{ https://doi.org/10.1103/PhysRevD.103.124001}{ Phys. Rev.D   \textbf{103}, 124001 (2022)}.

\bibitem{78}J.T.S.S. Junior, F.S.N. Lobo and M.E. Rodrigues, \emph{Black holes and regular black holes in coincident $f(Q,B_Q)$ gravity coupled to nonlinear electrodynamics}, \href{https://doi.org/10.1140/epjc/s10052-024-12696-8}{EPJC \textbf{84}, 332 (2024)}.

\bibitem{79} S. H. Strogatz,    \emph{Non Linear Dynamics and Chaos}, \href{https://www.biodyn.ro/course/literatura/Nonlinear_Dynamics_and_Chaos_2018_Steven_H._Strogatz.pdf}{ Published by CRC Press, Taylor and Francis Group, ISBN 13: 978-0-8133-4910-7,  \textbf{(2018)}}.

\bibitem{80}  M. Azreg-A¨ınou, S. Bahamonde and M. Jamil,   \emph{ Strong Gravitational Lensing by a Charged Kiselev Black Hole}, \href{https://doi.org/10.1140/epjc/s10052-017-4969-4}{EPJC \textbf{77}, 414 (2017)}.

\bibitem{81}  E. F. Eiroa and C. M. Sendra,   \emph{ Gravitational lensing by a regular black hole}, \href{}{Class. Quan. Grav    \textbf{28}, 085008 (2011)}.

\bibitem{82} S. Chandrasekhar,   \emph{ The Mathematical Theory of Black Holes}, \href{https://www.scribd.com/document/347738584/Chandrasekhar-S-the-Mathematical-Theory-of-Black-Holes}{Oxford Univ. Press, New York}.

\bibitem{83}] S. U. Islam, R. Kumar and S. G. Ghosh, \emph{ Gravitational lensing by black holes in the 4D Einstein-Gauss-Bonnet gravity}, \href{https://doi.org/10.1088/1475-7516/2020/09/030}{JCAP \textbf{09}, 030 (2020)}.

\bibitem{84} N.Tsukamotoand,   \emph{ Deflection angle in the strong deflection limit in a general asymptotically flat, static, spherically symmetric spacetime}, \href{ https://doi.org/10.1103/PhysRevD.95.064035}{Phys. Rev. D   \textbf{95}, 064035 (2017)}.

\bibitem{85} ] S. U. Islam, S. G. Ghosh and S. D. Maharaj,   \emph{Strong gravitational lensing by Bardeen black holes in 4D EGB gravity:constraints from supermassive black holes}, \href{https://doi.org/10.1016/j.cjph.2024.03.044}{Chi. Jour. Phys. \textbf{89}, 1710 (2024)}.

\bibitem{86}  Y. Wang et al., \emph{Strong Gravitational Lensing by Static Black Holes in Effective Quantum Gravity}, \href{https://doi.org/10.1140/epjc/s10052-025-13970-z}{EPJC    \textbf{85}, 302 (2025)}.

\bibitem{87} S. Weinberg,   \emph{ Gravitation and Cosmology: Principles and Applications of the General Theory of Relativity}, \href{https://www.scribd.com/document/375199795/Weinberg-S-Gravitation-and-Cosmology-Principles-and-Applications-of-the-General-Theory-of-Relativity-Wiley-1972-Isbn-0471925675-685S}{ John Wiley and Sons, New York    \textbf{(1972)}}.

\bibitem{88} S. Chen and J. Jing,    \emph{Strong Field Gravitational Lensing in the Deformed Hoˇrava-Lifshitz Black Hole}, \href{https://doi.org/10.1103/PhysRevD.80.024036}{Phys. Rev. D     \textbf{80}, 024036 (2009)}.

\bibitem{89} R. Zhang, J. Jing and S. Chen, \emph{ Strong gravitational lensing for black holes with scalar charge in massive gravity}, \href{ https://doi.org/10.1103/PhysRevD.95.064054}{Phys. Rev. D   \textbf{95}, 064054 (2017)}.

\bibitem{90} X. J. Gao et al.,    \emph{ Investigating strong gravitational lensing with black hole metrics modified with an additional term}, \href{https://www.sciencedirect.com/science/article/pii/S0370269321006237}{Phys. Lett. B \textbf{822}, 136683 (2021)}.

\bibitem{91} S. U. Islam, R. Kumar and S. G. Ghosh,  \emph{Gravitational lensing by black holes in the 4D Einstein-Gauss-Bonnet gravity}, \href{https://doi.org/10.1088/1475-7516/2020/09/030}{JCAP \textbf{09}, 030 (2020)}.

\bibitem{92} N. Tsukamotoand,    \emph{ Gravitational lensing in the Simpson-Visser black-bounce spacetime in a strong deflection limit}, \href{https://doi.org/10.1103/PhysRevD.103.024033}{Phys. Rev. D    \textbf{103}, 024033 (2021)}.

\bibitem{93} S. Chen and J. Jing,    \emph{Strong field gravitational lensing in the deformed Horava-Lifshitz black hole}, \href{https://doi.org/10.1103/PhysRevD.80.024036}{Phys. Rev. D \textbf{80}, 024036 (2009)}.

\bibitem{94}E. F. Eiroa and C. M. Sendra,    \emph{ Gravitational lensing by a regular black hole}, \href{https://doi.org/10.1088/0264-9381/28/8/085008}{Class. Quan. Grav.\textbf{28}, 085008 (2011)}.

\bibitem{95} C. Ding et al.,    \emph{ Strong Gravitational Lensing in a Noncommutative Black-Hole Spacetime}, \href{https://doi.org/10.1103/PhysRevD.83.084005}{Phys. Rev. D \textbf{83}, 084005 (2011)}.

\bibitem{96} J. S. U. Islam and S. G. Ghosh, \emph{Strong Gravitational Lensing by Loop Quantum Gravity Motivated Rotating Black Holes and EHT Observations}, \href{https://doi.org/10.1140/epjc/s10052-023-12205-3}{ EPJC \textbf{83}, 1014 (2023)}.

\bibitem{97} V. Bozza, S. Capozziello, G. Iovane and G. Scarpetta,    \emph{ Strong Field Limit of Black Hole Gravitational Lensing}, \href{https://doi.org/10.1023/A:1012292927358}{Gen. Rel. Grav    \textbf{33},1535 (2001)}.

\bibitem{98} S. Vagnozzi et al.,\emph{ Horizon-scale tests of gravity theories and
fundamental physics from the Event Horizon Telescope image
of Sagittarius A*}, \href{https://doi.org/10.1088/1361-6382/acd97b}{ Class. Quan Grav.    \textbf{40}, 165007 (2023)}.

\bibitem{99} T. Johannsen et al.,  \emph{ Testing general relativity with the shadow size of Sgr A*}, \href{https://doi.org/10.1103/PhysRevLett.116.031101}{ Phys. Rev. Lett. \textbf{116}, 031101 (2016)}.

\bibitem{100} D. Psaltis,   \emph{Testing general relativity with the Event Horizon Telescope}, \href{https://doi.org/10.1007/s10714-019-2611-5}{  Gen. Relativ. Gravit   \textbf{51}, 137 (2019)}.

\bibitem{101}  P. Kocherlakota et al.,   \emph{Constraints on black-hole charges with
the 2017 EHT observations of M87*}, \href{https://doi.org/10.1103/PhysRevD.103.104047}{Phys. Rev. D    \textbf{103}, 104047 (2021)}.

\bibitem{102}  R. Kumar and S.G. Ghosh, \emph{ Black hole parameter estimation from
its shadow}, \href{https://doi.org/10.3847/1538-4357/ab77b0}{Astrophys. J   \textbf{892}, 72 (2020)}.

\bibitem{103}  The EHT Collaboration,   \emph{ First Sagittarius A* Event Horizon Telescope Results. I. The Shadow of the Supermassive Black Hole in the Center of the Milky Way}, \href{https://doi.org/10.3847/2041-8213/ac6674}{Astrophy. J. Letts.   \textbf{930}, 21 (2022)}.

\bibitem{104}  Y. Dong,   \emph{The Gravitational Lensing by Rotating Black Holes in Loop Quantum Gravity}, \href{https://doi.org/10.1016/j.nuclphysb.2024.116612}{Nucl. Phys. B \textbf{1005}, 116612 (2024)}.

\bibitem{105}  X. M. Kuang et al.,   \emph{Constraining a Modified Gravity Theory in Strong Gravitational Lensing and Black Hole Shadow Observations}, \href{ https://doi.org/10.1103/PhysRevD.106.064012}{Phys. Rev. D \textbf{106}, 064012 (2022)}.





\end{thebibliography}
\end{document}